\documentclass[conference]{IEEEtran}
\IEEEoverridecommandlockouts

\usepackage{calc}
\usepackage{amsmath,amssymb,amsfonts}
\usepackage{algorithmic}
\usepackage{graphicx}
\usepackage{textcomp}
\usepackage{xcolor}
\usepackage{quantikz}
\usepackage{pdfcomment}
\usepackage{cleveref}
\usepackage{amsfonts} 
\usepackage{amsthm}
\usepackage{tabularx}
\usepackage{booktabs}
\usepackage{acro}
\usepackage{xspace}
\renewcommand{\ket}[1]{|#1\rangle}
\usepackage{todonotes}
\usepackage[all]{nowidow}
\usepackage{siunitx}
\usepackage{subcaption}
\usepackage{nicefrac}
\usepackage[inline]{enumitem}
\usepackage{balance}

\usepackage{tikz}
\usetikzlibrary{fit,calc,decorations,decorations.pathreplacing}

\usepackage[style=ieee, maxnames=9, minnames=6, date=year, doi=false, isbn=false, url=false, backend=biber, citestyle=numeric-comp, sortcites]{biblatex}
\AtEveryBibitem{
	\clearname{editor}
	\clearfield{series}
	\clearfield{isbn}
	\clearfield{issn}
	\clearfield{volume}
	\clearfield{number}
	\clearfield{pages}
	\clearfield{doi}
}

\hypersetup{
	pdflang={en-US},
	pdftitle={Automatic De-Quantization of Quantum Programs Using Constant Propagation},
	pdfauthor={Lian Remme, Alexander Weinert, Andre Waschk, Lukas Burgholzer, and Robert Wille},
	pdfsubject={Quantum computing},
	pdfkeywords={quantum-classical computing, optimization, quantum-classical optimization, constant propagation, compilation, formal semantics},
	draft={true}
}

\theoremstyle{definition}
\newtheorem{definition}{Definition}[section]

\DeclareAcronym{clops}{
	short=CLOPS,
	long=circuit layer operations per second,
}
\DeclareAcronym{mlir}{
	short=MLIR,
	long=Multi-Level Intermediate Representation,
}
\DeclareAcronym{nisq}{
	short=NISQ,
	long=Noisy Intermediate-Scale Quantum,
}
\DeclareAcronym{qft}{
	short=QFT,
	long=Quantum Fourier Transform,
}
\DeclareAcronym{qpe}{
	short=QPE,
	long=Quantum Phase Estimation,
}
\DeclareAcronym{qpu}{
	short=QPU,
	long=Quantum Processing Unit,
}

\def\BibTeX{{\rm B\kern-.05em{\sc i\kern-.025em b}\kern-.08em
    T\kern-.1667em\lower.7ex\hbox{E}\kern-.125emX}}

\makeatletter %
\newcommand{\linebreakand}{%
\end{@IEEEauthorhalign}
\hfill\mbox{}\par
\mbox{}\hfill\begin{@IEEEauthorhalign}
}
\makeatother %

\begin{document}

\title{Automatic De-Quantization of Quantum Programs Using Constant Propagation
}

\author{%
  \IEEEauthorblockN{%
  Lian Remme\IEEEauthorrefmark{1}\IEEEauthorrefmark{2}, %
  Alexander Weinert\IEEEauthorrefmark{1}, %
  Andre Waschk\IEEEauthorrefmark{1}, %
  Lukas Burgholzer\IEEEauthorrefmark{2}\IEEEauthorrefmark{3}, %
  Robert Wille\IEEEauthorrefmark{2}\IEEEauthorrefmark{3}}
\IEEEauthorblockA{%
  \IEEEauthorrefmark{1}Institute of Software Technology, German Aerospace Center (DLR), Cologne, Germany%
}
\IEEEauthorblockA{%
  \IEEEauthorrefmark{2}Chair for Design Automation, Technical University of Munich, Munich, Germany%
}
\IEEEauthorblockA{%
	\IEEEauthorrefmark{3}Munich Quantum Software Company GmbH, Garching near Munich, Germany %
}
\texttt{lian.remme@dlr.de}\qquad \texttt{alexander.weinert@dlr.de}\qquad \texttt{andre.waschk@dlr.de}\\
\texttt{lukas.burgholzer@tum.de}\qquad \texttt{robert.wille@tum.de}
}

\maketitle

\begin{abstract}
Quantum computing promises to solve problems beyond the reach of classical computers, but today’s quantum hardware is error-prone and much slower than classical hardware.
Every quantum operation is costly, making it crucial to minimize quantum resource usage in near-term algorithms.
Quantum resources should only be used when they are truly essential for quantum advantage, and not wasted on operations that can be efficiently handled by classical computation.

In this work, we focus on \emph{de-quantizing} quantum operations to classical computation whenever possible.
The approach we propose for this is \emph{hybrid quantum-classical constant propagation}, an optimization which reduces quantum operations by trading them for fast, reliable classical instructions.
This is done by tracking  between quantum and classical states to identify and eliminate unnecessary quantum gates and controls.

We formalize a hybrid state model for quantum-classical constant propagation, implement our optimizations in the open-source MQT Core tool, and evaluate them on benchmark circuits.

The obtained results show that quantum-classical constant propagation can reduce costly multi-qubit operations, making quantum programs more practical and robust for near-term devices.
This opens the door to new hybrid compiler strategies that leverage the best of both quantum and classical worlds.
\end{abstract}


\section{Introduction}

Quantum computing is a technology that uses the laws of quantum mechanics to process information.
Unlike classical bits, quantum bits (qubits) can exist in superpositions of zero and one.
Together with other properties like entanglement, this leads to quantum algorithms which perform certain computations faster than any known classical algorithm.

The quest for \emph{quantum advantage}, i.e., for solving problems that are intractable for classical computers~\cite{preskill2013quantum}, has driven advances in quantum algorithms and hardware.
Landmark algorithms such as Shor's prime factorization~\cite{shor1994algorithms} and Grover's search~\cite{grover1996fast} demonstrate the potential of quantum computation.

Yet, quantum calculations have a downside.
In the current \emph{\ac{nisq}} era, qubits are fragile, error-prone, and quantum operations are orders of magnitude slower than their classical counterparts.

To make quantum computing practical, we must use quantum resources only when they are essential for quantum advantage.
Whenever possible, we should \emph{de-quantize} operations to classical processors, to leveraging their speed and reliability as well as avoiding unnecessary quantum operations.

This principle applies at the level of algorithms and \emph{within} quantum calculations.
We ask:
Given a calculation, which parts \emph{must} remain quantum to unlock quantum advantage?
Which parts can efficiently be delegated to classical computation?
By answering these questions, we can accelerate quantum programs and make their execution more robust.

We propose \emph{quantum-classical constant propagation}: a static analysis that enables transformations which reduce quantum resource usage in quantum circuits.
These semantic-preserving transformations replace quantum operations with classical computation and control.
This is especially useful for multi-qubit gates, which are slow, error-prone, and challenging to implement on hardware with limited connectivity.

Our work expands the transformation routines available for quantum circuits.
We build on quantum constant propagation by Chen and Stade, which does not consider quantum-classical interplay~\cite{chen2023quantum}.
Similarly, our proposed Hadamard lifting extends measurement lifting by Rovara et al.~\cite{rovara2025qubit}.

We evaluate our transformations on benchmark circuits.
Our experiments show that quantum calculations can benefit from quantum-classical constant propagation.
We show that, in general, our quantum-classical approach reduces more quantum resources than the existing quantum constant propagation.

Some circuits benefit more from measurement lifting than from constant propagation.
Hadamard lifting, while promising, does not provide additional improvements.
We find that constant propagation and measurement lifting are complementary:
each excels in scenarios where the other is less effective, making the combination of both particularly powerful.

Our main contributions are:
\begin{itemize}
    \item Quantum-classical constant propagation: a technique for reducing quantum resources in hybrid quantum-classical computations.
    \item A formalization of the quantum-classical machine state, laying the foundation for constant propagation.
    \item An open-source implementation of constant propagation and Hadamard lifting in the Munich Quantum Toolkit (MQT) Core project~\cite{burgholzer2025MQTCore}\footnote{The functionality is given in a fork~\cite{mqt_core_fork}.}.
\end{itemize}

The remainder of the paper is organized as follows:
\Cref{sect:backgroundMotivation} provides background and motivating examples.
\Cref{sect:relatedwork} surveys related work.
\Cref{sec:hadamard-lifting} describes Hadamard lifting.
\Cref{sec:constant-propagation} formalizes our hybrid state model and complexity bounds and \Cref{sect:transformations-qcp} details the semantics-preserving transformations enabled by constant propagation.
Finally, \Cref{sect:experimental-evaluation} presents our evaluation, and \Cref{sect:conclusion} concludes with future directions.

\section{Motivation and Background}\label{sect:backgroundMotivation}

In this section, we review the necessary background knowledge for our work as well as the challenges in the intersection of quantum and classical computing.
We first give an overview over circuit-based quantum computing and the current state of the art of quantum hardware in \Cref{sect:backgroundMotivation:quantumComputing}.
Afterwards, we recall measurement lifting due to Rovara et al.~\cite{rovara2025qubit} and motivate our Hadamard lifting as an extension in \Cref{sect:backgroundMotivation:lifting}.
Finally, we recall quantum constant propagation due to Chen and Stade~\cite{chen2023quantum} and motivate our quantum-classical constant propagation as an extension in \Cref{sect:backgroundMotivation:constantPropagation}.

\subsection{Dynamic Circuit-based Quantum Computing}
\label{sect:backgroundMotivation:quantumComputing}

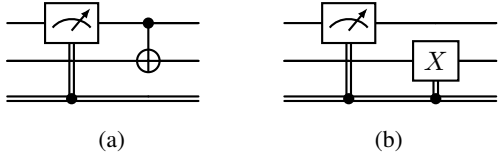
\begin{figure}[t]
	\centering
	\begin{subfigure}{.4\linewidth}
		\pdftooltip{
    	\begin{quantikz}[wire types={q,q,c},row sep={.5cm,between origins}]
			\lstick{} & \meter{} \wire[d][2]{c} & \ctrl{1} & \qw \\
			\lstick{} & \qw & \targ{}  & \qw\\
			\lstick{} & \control{} & &
		\end{quantikz}%
		}{A quantum circuit with two qubits and one bit. The applied gates are (in order): The first qubit is measured, the result is written into the bit. A CNOT is applied with the first qubit as control and the second as target.}
		\caption{}
		\label{subfig:dynamicCircuitExampleA}
	\end{subfigure}
	\begin{subfigure}{.4\linewidth}
		\pdftooltip{
    	\begin{quantikz}[wire types={q,q,c},row sep={.5cm,between origins}]
			\lstick{} & \meter{} \wire[d][2]{c} & \qw & \qw \\
			\lstick{} & \qw & \gate{X}  & \qw\\
			\lstick{} & \control{} & \control{} \wire[u][1]{c} &
		\end{quantikz}%
		}{A quantum circuit with two qubits and one bit. The applied gates are (in order): The first qubit is measured, the result is written into the bit. A NOT controlled classically by the first bit is applied to the second qubit.}
		\caption{}
		\label{subfig:dynamicCircuitExampleB}
	\end{subfigure}
	\caption{An example for a dynamic circuit. The left and right circuits are equivalent. The quantum control is removed in exchange for a classical condition on a gate.}
	\label{fig:dynamicCircuitExample}
\end{figure}

Quantum circuits and their typical graphical notation are extensively used in this work.
We refer the reader to~\cite[Chapter~4]{nielsen2010} for an introduction to these topics.

Currently available quantum computers are error-prone.
Operations on IBM's superconducting \acp{qpu} suffer error rates between $10^{-2}$ and $10^{-4}$, with measurement being especially error-prone.
Only single-qubit gates offer relative robustness~\cite{ibm_aachen_infos,ibm_boston_infos,ibm_kingston_infos}.
IonQ's \ac{qpu} Aria shows similar error rates (between $10^{-3}$ and $10^{-4}$~\cite{ionq_aria}).
In contrast, the error rates of classical CPUs are negligible.

Quantum and classical computers also differ critically in their execution speed.
IBM's superconducting \acp{qpu} execute a few hundred thousand layer operations per second~\cite{ibm_aachen_infos,ibm_boston_infos,ibm_kingston_infos}.
Executing a single instruction on IonQ's Aria~\cite{ionq_aria} can take hundreds of microseconds (\SI{135}{\micro\second} for one-qubit operations, \SI{600}{\micro\second} for two-qubit gate operations).
This translates to just a few thousand quantum operations per second.
Meanwhile, classical CPUs execute billions of instructions per second~\cite{intel_cpu_infos}.

A typical approach to reduce the errors of faulty quantum resources is the use of \emph{dynamic circuits}, i.e., hybrid circuits that use gates guarded by registers.
Dynamic circuits can use measurement outcomes to control subsequent quantum operations or to trigger classical computation and branching.
Thus, they unlock new possibilities for efficiency.
For instance, when a quantum control immediately follows a measurement, it can be replaced with a classical condition, eliminating costly multi-qubit gates.
We illustrate this simplification in \Cref{subfig:dynamicCircuitExampleA} and \Cref{subfig:dynamicCircuitExampleB}.
This strategy has been applied manually for practical gains:
Algorithms like iterative phase estimation~\cite{dobvsivcek2007arbitrary} and Shor's algorithm for $2n+1$ qubits~\cite{beauregard2002circuit} exploit dynamic circuits to minimize quantum resource usage.

Since currently available quantum computers are severely limited in terms of their available quantum resources, the optimization of quantum circuits is an active area of research~\cite{rattacaso2025quantum,quetschlich2023compiler,zen2025reusability,dang2025qukkos,karuppasamy2025comprehensive,rovara2025qubit,praba2025integration,sunkel2025quantum,li2024quarl,fosel2021quantum,arora2025local,forster2025quantum,hopf2026quantum,quetschlich2023predicting,quetschlich2024towards}.
One goal of the optimizations is to reduce quantum resources, e.\,g. the number of qubits~\cite{karuppasamy2025comprehensive,rovara2025qubit}, circuit depth~\cite{karuppasamy2025comprehensive,praba2025integration,sunkel2025quantum,fosel2021quantum}, or gate count~\cite{karuppasamy2025comprehensive,li2024quarl,fosel2021quantum,arora2025local,forster2025quantum}.
This is usually done by removing unnecessary quantum operations.
However, another approach is to \emph{replace} quantum operations with classical operations whenever possible.

In the following subsections, we present two particular optimizations, namely quantum constant propagation by Chen and Stade~\cite{chen2023quantum} and measurement lifting by Rovara et al.~\cite{rovara2025qubit}.
We moreover motivate our respective extensions, namely quantum-classical constant propagation and Hadamard lifting.

\subsection{Measurement Lifting}
\label{sect:backgroundMotivation:lifting}

Measurement lifting was introduced by Rovara et al.~\cite{rovara2025qubit} as part of their qubit-reuse strategy.
It is an optimization that pushes measurements as far forward in the circuit as possible.
This exposes opportunities to simplify or eliminate quantum operations, making circuits more efficient and robust.

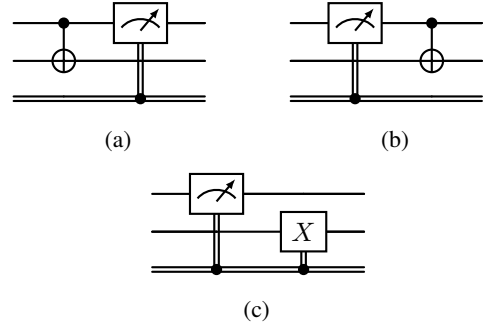
\begin{figure}[t]
	\centering
	\begin{subfigure}{.4\linewidth}
		\pdftooltip{
    	\begin{quantikz}[wire types={q,q,c},row sep={.5cm,between origins}]
			\lstick{} & \ctrl{1} & \meter{} \wire[d][2]{c} & \qw \\
			\lstick{} & \targ{} & \qw & \qw\\
			\lstick{} & & \control{} &
		\end{quantikz}%
		}{A quantum circuit with two qubits and one bit. The applied gates are (in order): A CNOT is applied with the first qubit as control and the second as target. The first qubit is measured, the result is written into the bit.}
		\caption{}
		\label{fig:measurementLiftingRoutineA}
	\end{subfigure}
	\begin{subfigure}{.4\linewidth}
		\pdftooltip{
    	\begin{quantikz}[wire types={q,q,c},row sep={.5cm,between origins}]
			\lstick{} & \meter{} \wire[d][2]{c} & \ctrl{1} & \qw \\
			\lstick{} & \qw & \targ{} & \qw\\
			\lstick{} & \control{} & &
		\end{quantikz}%
		}{A quantum circuit with two qubits and one bit. The applied gates are (in order): The first qubit is measured, the result is written into the bit. A CNOT is applied with the first qubit as control and the second as target.}
		\caption{}
		\label{fig:measurementLiftingRoutineB}
	\end{subfigure}
	\begin{subfigure}{.4\linewidth}
		\pdftooltip{
    	\begin{quantikz}[wire types={q,q,c},row sep={.5cm,between origins}]
			\lstick{} & \meter{} \wire[d][2]{c} & \qw & \qw \\
			\lstick{} & \qw & \gate{X} & \qw\\
			\lstick{} & \control{} & \control{} \wire[u][1]{c} &
		\end{quantikz}%
		}{A quantum circuit with two qubits and one bit. The applied gates are (in order): The first qubit is measured, the result is written into the bit. A NOT controlled classically by the first bit is applied to the second qubit.}
		\caption{}
		\label{fig:measurementLiftingRoutineC}
	\end{subfigure}
	\caption{Measurement lifting in action: By commuting the measurement from \Cref{fig:measurementLiftingRoutineA} forward to create the circuit in \Cref{fig:measurementLiftingRoutineB}, a quantum control can be replaced with a classical control. This creates the circuit in \Cref{fig:measurementLiftingRoutineC} and eliminates a costly two-qubit gate. All three circuits are equivalent.}
	\label{fig:measurementLiftingRoutine}
\end{figure}

The core idea is that by commuting measurements past certain gates, we can often replace quantum controls with classical ones, or remove unnecessary gates altogether.
We illustrate this in \Cref{fig:measurementLiftingRoutine}, where we commute a measurement past a quantum control, i.e., we move from \Cref{fig:measurementLiftingRoutineA} to \Cref{fig:measurementLiftingRoutineB}.
Afterwards, the quantum control can be replaced by a classical control, seen in \Cref{fig:measurementLiftingRoutineC}.
This reduces the number of expensive two-qubit gates.

Measurements can be commuted with phase gates since they do not affect measurement outcomes, and with Pauli X and Y gates if the measurement outcome is negated.
Measurement lifting can reduce quantum control and create gates that do not influence subsequent measurements.
These gates can then be trivially removed, decreasing the depth of the circuit.

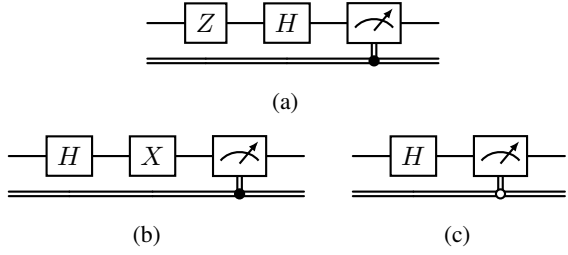
\begin{figure}[t]
	\centering
	\begin{subfigure}{.5\linewidth}
		\pdftooltip{
     	\begin{quantikz}[wire types={q,c},row sep={.5cm,between origins}]
			\lstick{} & \gate{Z} & \gate{H} & \meter{} \wire[d][1]{c} &\\
			& & & \control{} &
		\end{quantikz}%
		}{A quantum circuit with one qubit and one bit. The applied gates are (in order): A Pauli-Z gate to the qubit. A Hadamard gate to the qubit. The qubit is measured, the result is written into the bit.}
		\caption{}
		\label{fig:HLiftingExampleA}
	\end{subfigure}
	\begin{subfigure}{.5\linewidth}
		\pdftooltip{
    	\begin{quantikz}[wire types={q,c},row sep={.5cm,between origins}]
			\lstick{} & \gate{H} & \gate{X} & \meter{} \wire[d][1]{c} &\\
			& & & \control{} &
		\end{quantikz}%
		}{A quantum circuit with one qubit and one bit. The applied gates are (in order): A Hadamard gate to the qubit. A Pauli-X gate to the qubit. The qubit is measured, the result is written into the bit.}
		\caption{}
		\label{fig:HLiftingExampleB}
	\end{subfigure}
	\begin{subfigure}{.4\linewidth}
		\pdftooltip{
    	\begin{quantikz}[wire types={q,c},row sep={.5cm,between origins}]
			\lstick{} & \gate{H} & \meter{} \wire[d][1]{c} &\\
			& & \control[open]{} &
		\end{quantikz}%
		}{A quantum circuit with one qubit and one bit. The applied gates are (in order): A Hadamard gate to the qubit. The qubit is measured, the result negated and written into the bit.}
		\caption{}
		\label{fig:HLiftingExampleC}
	\end{subfigure}
	\caption{Hadamard lifting: By commuting Pauli and Hadamard gates, measurements can be lifted further, enabling more quantum resource reduction. To create the circuit in \Cref{fig:HLiftingExampleB} from \Cref{fig:HLiftingExampleA}, the Hadamard gate is lifted in front of the Pauli gate. Afterwards, the measurement is lifted in front of the Pauli gate and the result negated, to create \Cref{fig:HLiftingExampleC}. All three circuits are equivalent.}
	\label{fig:HLiftingExample}
\end{figure}

Measurement lifting can be blocked by Hadamard gates before a measurement.
As Hadamard gates put qubits into a superposition, they prevent measurements from being lifted.
To overcome this, we introduce \emph{Hadamard lifting}: a technique that leverages the commutation relations between Hadamard and Pauli gates to move Hadamards away from measurements.

For example, a Hadamard gate followed by a Pauli-Z gate is equivalent to a Pauli-X gate followed by a Hadamard gate.
By applying these commutation rules we can transform circuits so that measurements can be lifted even further, unlocking new opportunities for classical control and further quantum resource reduction.

We illustrate this in \Cref{fig:HLiftingExample}, where we first commute the Hadamard gate in front of the Pauli gate, obtaining the circuit in \Cref{fig:HLiftingExampleB} from that in \Cref{fig:HLiftingExampleA}.
Afterward, we can lift the measurement in front of the Pauli gate to obtain the circuit in \Cref{fig:HLiftingExampleC}.

\subsection{Quantum Constant Propagation}
\label{sect:backgroundMotivation:constantPropagation}

\begin{figure}[t]
	\centering
	\begin{subfigure}{.45\linewidth}
		\pdftooltip{
		\parbox{.95\linewidth} {%
    	\begin{quantikz}[wire types={q,q,q,c},row sep={.5cm,between origins}]
			\lstick{} & \ctrl{1} & \meter{} \wire[d][3]{c} & \qw & \qw\\
			\lstick{} & \targ{} & \qw & \ctrl{1} & \qw\\
			\lstick{} & \qw & \qw & \targ{} & \qw\\
			\lstick{} & & \control{} & &
		\end{quantikz}%
		}
		}{A quantum circuit with three qubits and one bit. The applied gates are (in order): A CNOT gate with the first qubit as control and second as target. The first qubit is measured, the result is written into the bit. A CNOT gate with the second qubit as control and third as target.}
		\caption{}
		\label{fig:hcpExampleA}
	\end{subfigure}
	\begin{subfigure}{.45\linewidth}
		\pdftooltip{
		\parbox{.95\linewidth} {%
    	\begin{quantikz}[wire types={q,q,q,c},row sep={.5cm,between origins}]
			\lstick{} & \ctrl{1} & \meter{} \wire[d][3]{c} & \qw & \qw\\
			\lstick{} & \targ{} & \qw & \qw & \qw\\
			\lstick{} & \qw & \qw & \gate{X} & \qw\\
			\lstick{} & & \control{} & \control{} \wire[u][1]{c} &
		\end{quantikz}%
		}
		}{A quantum circuit with three qubits and one bit. The applied gates are (in order): A CNOT gate with the first qubit as control and second as target. The first qubit is measured, the result is written into the bit. A NOT controlled classically by the first bit is applied to the third qubit.}
		\caption{}
		\label{fig:hcpExampleB}
	\end{subfigure}
	\caption{A circuit for which constant propagation is useful: the state of the second qubit becomes determined by a prior measurement in \Cref{fig:hcpExampleA}, allowing a quantum control to be replaced by a classical condition in \Cref{fig:hcpExampleB}. Both circuits are equivalent.}
	\label{fig:hcpExample}
\end{figure}
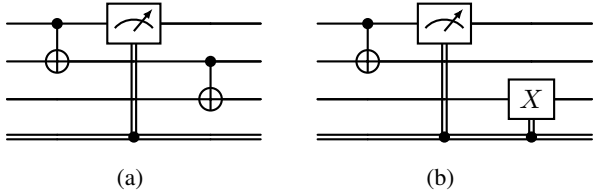

Hadamard and measurement lifting are pattern-based optimization routines.
While this makes the algorithms straightforward to run, it reduces the cases which they can be applied to.
For example, the circuit in \Cref{fig:hcpExampleA} would not be improved by measurement lifting, even though a transformation to the circuit in \Cref{fig:hcpExampleB} would be possible.

We could aim to find a comprehensive catalog of patterns for replacing quantum control by classical control.
However, we aim for a more general solution: constant propagation.
We simulate a restricted, efficiently representable abstraction of the quantum-classical machine state.
The analysis detects relations between qubits and registers that pattern matching misses. 
Using these relations, we can then replace quantum control by classical control.

Constant propagation is also used for classical programs.
It propagates the data-flow of a program to check whether a variable has a fixed value at a certain point~\cite[Chapter 9.4]{aho2007compilers}.

Chen and Stade show how to propagate quantum-state information through a circuit to find superfluous gates~\cite{chen2023quantum}.
If a control qubit is always $0$, the controlled operation can be removed.
If a control qubit is always $1$, the control can be removed.
Additionally, if a gate is controlled by more than one qubit, it is checked whether all its controls can be simultaneously satisfied.
If not, the gate is removed.
Also, if more than one qubit controls an operation, it is checked whether one or more controlling qubits are implied by others.
In that case, the implied qubit(s) can be removed.

We extend this idea to dynamic circuits: 
Our quantum machine abstraction tracks both quantum amplitudes and correlated registers.
This enables transformations as depicted in \Cref{fig:hcpExample}.

\section{Related Work}\label{sect:relatedwork}

Measurements can provide different functionality in quantum circuits.
Uotila et al.~\cite{uotila2025perspectives} evaluated the role of measurements in quantum circuits.
They distinguish three different usages:
Read-out at the end of static circuits, read-out in dynamic circuits and succeeding modifications depending on the read-out results, and measurements to do quantum error correction.
These different usages of measurements emphasize how little static circuits utilize measurements.
Our work modifies circuits such that measurements that used to be read-out at the end of static circuits become read-out in dynamic circuits.

There are several works about quantum circuit optimization.
A survey by Karuppasamy et al.~\cite{karuppasamy2025comprehensive} classifies optimization techniques.
There are heuristic ones which find effective solutions or approximations.
Heuristics are often applied iteratively and swap, merge or decompose gates.
Machine learning methods utilize machine learning for optimization.
Unitary synthesis works with unitary matrices which represent quantum circuits.
Circuits are supposed to be efficiently constructed and factorized from the matrices.
The last technique is the algorithmic approach.
In this case, systematic strategies are applied.
The survey considers hybrid quantum classical algorithms, but not dynamic circuits or their optimization.

Another survey by Yan et al.~\cite{yan2024quantum} points out the increasing relevance of artificial intelligence in circuit optimization.
They also contribute a website with a list of circuit synthesis or compilation strategies that use AI.

Arora et al.~\cite{arora2025local} contributed an optimization routine to reduce quantum gates.
The routine locally optimizes sub-circuits and merges the sub-circuits later on.
Hao, Xu and Tannu~\cite{hao2025reducing} provided unitary synthesis to reduce T count, Clifford gate count, and approximation errors.
A hybrid evolutionary algorithm by Sünkel et al.~\cite{sunkel2025quantum} is used for circuit depth minimization.
Rattacaso et al.'s optimization routine~\cite{rattacaso2025quantum} optimizes circuits during quantum compilation routines.

Besides these routines are also some which use machine learning approaches.
Praba et al.~\cite{praba2025integration} worked on both supervised and reinforcement learning strategies.
Quetschlich, Burgholzer and Wille~\cite{quetschlich2023compiler}, as well as Mills et al.~\cite{mills2026reinforcement}, implemented reinforcement learning agents which decide on the optimize application order.
Alpha-Tensor quantum is a reinforcement learning agent to reduce T gate count by Zen, Nägele and Marquardt~\cite{zen2025reusability}.
Another reinforcement learning tool for circuit optimization, Quarl, has been introduced by Li, Ding und Xie~\cite{li2024quarl}.

All of the mentioned works work on static quantum circuits.
This means that none of them utilized the possibility to use classical resources instead of quantum ones.
In contrast, the following papers address work on dynamic circuits in the context of circuit optimization:

Remme, Weinert and Waschk evaluated hybrid optimization routines and how to consider classical computing resources within dynamic circuits~\cite{remme2025optimization}.
Ye et al.~\cite{ye2026qspe} provided the Quantum Skeletal Program Enumeration (QSPE), a tool to validate and test compilers that work on dynamic circuits.
They also created a set of dynamic circuits as benchmarks.
Qukkos by Dang, Son and Kim~\cite{dang2025qukkos} uses measurement-based quantum computing~\cite{briegel2009measurement} to lower circuit depth and improve fidelity. 
Quantum circuits are transformed into measurement operations with adaptive corrections.
The exact transformations are not given in the paper.

None of these works use classical calculations to reduce quantum resources.
Our work fills this gap.

\section{Hadamard Lifting}\label{sec:hadamard-lifting}

In this section, we aim to improve the efficiency of measurement lifting~\cite{rovara2025qubit}.
We introduce Hadamard lifting: A gate re-ordering routine to improve the results of measurement lifting.
This could provide additional effectiveness.

\subsection{Lifting General Pauli-Gates Over Hadamard Gates}

Measurement lifting~\cite{rovara2025qubit} is an optimization routine which lifts measurements in front of certain gates.
This is done to decrease the number of gates that are needed.

While measurements can be lifted over Pauli- and phase-gates, they cannot be lifted over Hadamard gates.
To increase the efficiency of measurement lifting, we use the following commutation properties of Hadamard gates to move them away from measurements:

\begin{center}
	\pdftooltip{
  	\begin{quantikz}[wire types={q},row sep={.5cm,between origins}]
		\lstick{} & \gate{X} & \gate{H} & \qw
	\end{quantikz}
	$\approx$
	\begin{quantikz}[wire types={q},row sep={.5cm,between origins}]
		\lstick{} & \gate{H} & \gate{Z} & \qw
	\end{quantikz}
	}{Two equivalent quantum circuits: Both show one qubit. On the first, first a Pauli-X gate and then a Hadamard gate are applied. On the second, first a Hadamard gate and then a Pauli-Z gate are applied.},\\
	\pdftooltip{
	\begin{quantikz}[wire types={q},row sep={.5cm,between origins}]
		\lstick{} & \gate{Z} & \gate{H} & \qw
	\end{quantikz}
	$\approx$
	\begin{quantikz}[wire types={q},row sep={.5cm,between origins}]
		\lstick{} & \gate{H} & \gate{X} & \qw
	\end{quantikz}
	}{Two equivalent quantum circuits: Both show one qubit. On the first, first a Pauli-Z gate and then a Hadamard gate are applied. On the second, first a Hadamard gate and then a Pauli-X gate are applied.},\\
	\pdftooltip{
	\begin{quantikz}[wire types={q},row sep={.5cm,between origins}]
		\lstick{} & \gate{Y} & \gate{H} & \qw
	\end{quantikz}
	$\approx$
	\begin{quantikz}[wire types={q},row sep={.5cm,between origins}]
		\lstick{} & \gate{H} & \gate{Y} & \qw
	\end{quantikz}
	}{Two equivalent quantum circuits: Both show one qubit. On the first, first a Pauli-Y gate and then a Hadamard gate are applied. On the second, first a Hadamard gate and then a Pauli-Y gate are applied.},
\end{center}
where we write~$\approx$ to denote semantical equivalence of quantum circuits.

In case of commuting a Pauli-Y and a Hadamard gate, the circuit receives a global phase of $-1$, which we account for by adding a global phase gate to the circuit.

These commutations are also applicable if the Hadamard- and Pauli-X or Pauli-Z gate are controlled by exactly the same qubits.
In this case, there may not be any other gate applied between the controlled gates.
In case of a controlled Pauli-Y gate, a relative phase would be added by commuting the gates.

\subsection{Lifting Controlled Pauli-Z-Gates Over Hadamard Gates}

\begin{figure}[th]
	\centering
	\pdftooltip{
  	\begin{quantikz}[row sep={.5cm,between origins}]
    	\lstick{} & \ctrl{1} & \gate{H} & \qw \\
    	\lstick{} & \gate{Z}  & \ctrl{-1} & \qw
  	\end{quantikz}%
  	}{A quantum circuit with two qubits. The applied gates are (in order): A controlled Pauli-Z gate with the first qubit as control and second as target. A controlled Hadamard gate with the second qubit as control and first as target.}
  	$\approx$
  	\pdftooltip{
  	\begin{quantikz}[row sep={.5cm,between origins}]
    	\lstick{} & \gate{Z} & \gate{H} & \qw \\
    	\lstick{} & \ctrl{-1}  & \ctrl{-1} & \ghost{Z} \qw
  	\end{quantikz}%
  	}{A quantum circuit with two qubits. The applied gates are (in order): A controlled Pauli-Z gate with the second qubit as control and first as target. A controlled Hadamard gate with the second qubit as control and first as target.}\\
  	$\approx$
  	\pdftooltip{
  	\begin{quantikz}[row sep={.5cm,between origins}]
    	\lstick{} & \gate{H} & \gate{X} & \qw \\
    	\lstick{} & \ctrl{-1}  & \ctrl{-1} & \ghost{Z}  \qw
  	\end{quantikz}%
  	}{A quantum circuit with two qubits. The applied gates are (in order): A controlled Hadamard gate with the second qubit as control and first as target. A controlled Pauli-X gate with the second qubit as control and first as target.}
	\caption{%
    We can commute a controlled Pauli-Z gate with a controlled Hadamard even if the targets are not applied to the same qubit. %
    For this we use the fact that the Pauli-Z gate can be applied to any positively controlled control instead of the original target, meaning we can transform the first circuit into the second one. %
    Afterward, the Hadamard- and Pauli-gate can be commuted, resulting in the third circuit. %
  }
	\label{fig:ctrl-pauli-commuting}
\end{figure}
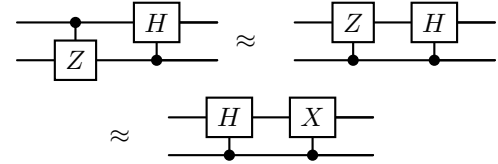

If we have positively controlled Pauli-Z gates, it is not relevant which qubit is the target and which one is the control.
Therefore, we can apply the routine depicted in \Cref{fig:ctrl-pauli-commuting} to positively controlled Pauli-Z gates and Hadamard gates which are applied to the same qubits, even if they do not have the same target qubit.

\subsection{Commuting a Single Hadamard Through CNOT}

For CNOT gates, the following equivalence holds:

\begin{center}
	\pdftooltip{
  	\begin{quantikz}[row sep={.75cm,between origins}]
		\lstick{} & \ctrl{1} & \qw \\
		\lstick{} & \targ{}  & \qw
	\end{quantikz}
	}{A quantum circuit with two qubits. The applied gates are (in order): A CNOT gate with the first qubit as control and second as target.}
	$\approx$
	\pdftooltip{
  	\begin{quantikz}[row sep={.75cm,between origins}]
		\lstick{} & \gate{H} & \targ{} & \gate{H} & \qw \\
		\lstick{} & \gate{H} & \ctrl{-1}  & \gate{H} & \qw
	\end{quantikz}
	}{A quantum circuit with two qubits. The applied gates are (in order): A Hadamard gate applied to the first and a Hadamard gate applied to the second qubit. A CNOT gate with the second qubit as control and first as target. A Hadamard gate applied to the first and a Hadamard gate applied to the second qubit.}
\end{center}

\begin{figure}[th]
	\centering
	\pdftooltip{
    \begin{quantikz}[row sep={.75cm,between origins},column sep={.3cm}]
		\lstick{} & \ctrl{1} & \qw & \qw \\
		\lstick{} & \targ{}  & \gate{H} & \meter{}
	\end{quantikz}%
	}{A quantum circuit with two qubits. The applied gates are (in order): A CNOT gate with the second qubit as control and first as target. A Hadamard gate applied to the second qubit. A measurement of the second qubit.}
    \hfill$\approx$\hfill
    \pdftooltip{
    \begin{quantikz}[row sep={.75cm,between origins},column sep={.3cm}]
		\lstick{} & \gate{H} & \targ{} & \gate{H} & \qw & \qw \\
		\lstick{} & \gate {H} & \ctrl{-1}  & \gate{H} & \gate{H} & \meter{}
	\end{quantikz}%
	}{A quantum circuit with two qubits. The applied gates are (in order): A Hadamard gate applied to the first and a Hadamard gate applied to the second qubit. A CNOT gate with the second qubit as control and first as target. A Hadamard gate applied to the first and a Hadamard gate applied to the second qubit. A Hadamard gate applied to the second qubit. A measurement of the second qubit.}
    \\ $\approx$
    \pdftooltip{
    \begin{quantikz}[row sep={.75cm,between origins},column sep={.3cm}]
		\lstick{} & \gate{H} & \targ{} & \gate{H} & \qw \\
		\lstick{} & \gate {H} & \ctrl{-1} & \meter{}
	\end{quantikz}%
	}{A quantum circuit with two qubits. The applied gates are (in order): A Hadamard gate applied to the first and a Hadamard gate applied to the second qubit. A CNOT gate with the second qubit as control and first as target. A Hadamard gate applied to the first qubit. A measurement of the second qubit.}
	\caption{%
    If we have a CNOT target followed by a Hadamard gate followed by a measurement, one can change this to a measurement following the control. %
    First, the target and control is flipped by applying four new Hadamard gates, which changes the circuit from the first to the second one. %
    Then, the two Hadamard gates after each other cancel out, yielding the third circuit. %
  }
	\label{fig:cnot-commuting}
\end{figure}
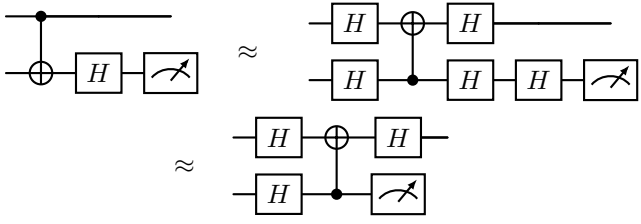

Therefore, if we have a Hadamard gate between a CNOT target and a measurement, we can do the transformation as depicted in \Cref{fig:cnot-commuting}.
This is useful for measurement lifting.
After the transformation, the measurement can be lifted before the control, and the result of the measurement can be used as classical control, while the quantum control can be removed.

This routine introduces more gates to the circuit than it removes.
However, as stated in \Cref{sect:backgroundMotivation:quantumComputing}, multi-qubit gates take much more time and are much more error-prone than single-qubit gates.
That justifies this transformation.

\section{Constant Propagation}\label{sec:constant-propagation}

In this section, we will explain quantum-classical constant propagation and define the abstract state we use for this.
Measurement lifting is applied to simplify the circuit close to measurements.
Meanwhile, constant propagation is most effective at the start of a circuit.

We first recall some basic notation we require for this work.

\newcommand{\bools}{\mathbb B}
\newcommand{\nats}{\mathbb N}
\newcommand{\complex}{\mathbb C}
\newcommand{\set}[1]{\{ #1 \}}
\newcommand{\card}[1]{| #1 |}
\newcommand{\abs}[1]{|\,#1\,|}
\newcommand{\norm}[1]{||\,#1\,||}

We define~$\bools = \set{0,1}$, we write~$\nats$ to denote the positive natural numbers, and we write~$\complex$ to denote the complex numbers.
Moreover, for~$n \in \nats$ we define $[n] = \set{1,\dots,n}$ and, given a sequence~$\beta = b_1\cdots b_n$ we define~$\beta[i \rightarrow x] = b_1 \cdots b_{i-1} x b_{i+1} \cdots b_n$.

\subsection{Concrete Execution}
\label{sec:constant-propagation:concrete}

We begin by formalizing the semantics of executing a quantum circuit.
In this section, we provide a denotational semantics of quantum circuits, which we view as a series of instructions.
Each instruction is either a guarded or unguarded application of some gate to the qubits, a measurement of a qubit, or a reset of a qubit.
We do not consider the parallelism inherent in the circuit notation and instead identify a circuit with an arbitrary, semantic-preserving interleaving of its instructions.

For the remainder of this work, fix some number~$n \in \nats$ of qubits.
A \emph{(concrete) quantum state}~$\ket{\psi}$ is a mapping~$\ket{\psi}\colon \bools^n \rightarrow \complex$ that satisfies $\left(\Sigma_{q \in \bools^n} \card{\ket{\psi}(q)}^2 = 1\right)$.
We write~$\Psi$ to denote the set of all concrete quantum states.
Given a mapping $f\colon \bools^n \rightarrow \complex$ we define its normalization $\norm{f}(q) = f(q) / (\Sigma_{q \in \bools^n} \card{f(q)}^2)$.

For the remainder of this work, fix some number~$m \in \nats$ of registers.
A \emph{(concrete) hybrid state} $\varphi = (\ket{\psi}, \beta)$ comprises a quantum state~$\ket{\psi}$ and a bit sequence $\beta \in \bools^m$.
We write~$\Phi$ to denote the set of all hybrid states.
A \emph{(concrete) machine state}~$\rho$ is a probability distribution over~$\Phi$, i.e., we have $\rho \colon \Phi \rightarrow [0; 1]$ and $(\sum_{\varphi \in \Phi} \rho(\varphi)) = 1$.

\newcommand{\concapp}[1]{[\![ #1 ]\!]}

We now define the effects of operations on machine states.
We formalize these by defining a family of functions~$\concapp{\cdot}$ for the unguarded and guarded applications of unitary gates, as well as measurements and resets of qubits.
The function~$\concapp{\cdot}$ takes a machine state as an input and yields a resulting machine state.

We first define the effect of applying a gate to a hybrid state.
Let~$\varphi = (\ket\psi, \beta)$ be a hybrid state and pick $\ket{\psi'}$ such that $U\ket{\psi'} = \ket\psi$.
The probability of the machine being in state~$\varphi$ after applying~$U$ equals the probability of it being in state $\ket{\psi'}$ before applying~$U$.
As gate applications are reversible,~$\ket{\psi'}$ is unique and we arrive at the following definition.

\newcommand{\app}{\texttt{\textbf{app}}}

\begin{definition}[$\concapp{\app\ U}$]
  Let~$U$ be a unitary gate.
  We define the effect of applying~$U$ to a machine state~$\rho$ as
  \[ \concapp{\app\ U}(\rho) \colon (\ket{\psi}, \beta) \mapsto \rho(U^{-1}\ket{\psi}, \beta) \]
\end{definition}

\newcommand{\capp}{\texttt{\textbf{capp}}}

Gates can also be guarded by registers:
A guarded gate~$U$ is applied only if certain registers are set to~$1$.
W.l.o.g, we treat only this case in this work, and omit the dual case where~$U$ is applied if certain registers are set to~$0$.

\begin{definition}[$\concapp{\capp\ U\ i_1 \dots i_k}$]
  Let~$U$ be a unitary gate and let~$i_1 < \cdots < i_k \in [n]$.
  We define the effect of applying~$U$ guarded by the registers~$i_1,\dots,i_k$ to the machine state~$\rho$ as
  \begin{multline*}
  	\concapp{\capp\ U\ i_1 \dots i_k}(\rho) \colon (\ket{\psi}, \beta) \mapsto \\
  	\begin{cases}
      \rho(U^{-1}\ket{\psi}, \beta) & \text{if } \forall j \in [k] : b_j = 1 \\
  		\rho(\ket{\psi}, \beta) & \text{otherwise} \\
  	\end{cases}
  \end{multline*}
  where $\beta = b_1 \cdots b_m$.
\end{definition}

\newcommand{\meas}{\textbf{\texttt{meas}}}
\newcommand{\reset}{\textbf{\texttt{reset}}}
\newcommand{\mpred}{\text{mpred}}
\newcommand{\rpred}{\text{rpred}}

We now proceed to define the result of measuring qubits.
Recall that measuring a qubit not only causes the state of the measured qubit to collapse, but also influences all qubits that are entangled with the measured one.
Hence, we first define the effect of measuring on a quantum state.

\newcommand{\Prob}{\text{Prob}}

\begin{definition}[Measurement]
  Let $\ket\psi$ be a quantum state, let~$b \in \bools$ and let~$i \in [n]$.
We define the quantum state resulting from measuring the $i$-th qubit as value~$b$ as $\ket\psi_{i \rightarrow b} = \norm{f}$ where $f \colon \bools^n \rightarrow \complex$ with \[
  f(q_1 \cdots q_n) = \begin{cases}
    \ket\psi(q_1 \cdots q_n) & \text{if $q_i = b$}\\
    0 & \text{otherwise}
  \end{cases}
\]
We moreover define the probability of measuring the~$i$-th qubit as value~$b$ as
\[ \Prob_{\ket\psi}(i \rightarrow b) = \Sigma_{q = q_1 \cdots q_n \in \bools^n, q_i = b} \abs{\ket\psi(q)}^2 \enspace . \]
\end{definition}

Intuitively, we say that a hybrid state~$\varphi = (\ket\psi, \beta)$ is a $i \rightarrow j$-predecessor of a hybrid state~$\varphi' = (\ket{\psi'}, \beta')$ with~$\beta' = b'_1 \cdots b'_m$ if~$\beta$ and~$\beta'$ differ at most position~$j$ and if $\ket{\psi'}$ results from~$\ket\psi$ by measuring the $i$-th qubit as~$b'_j$.

\begin{definition}[$i \rightarrow j$-predecessors]
Let $\varphi = (\ket\psi, \beta)$ with $\beta = b_1 \cdots b_m$ and $\varphi' = (\ket{\psi'}, \beta')$ with $\beta'= b'_1 \cdots b'_m$ be hybrid states.
We say that $\varphi$ is a \emph{$i \rightarrow j$-predecessor} of $\varphi'$ if $\beta' = \beta[j \rightarrow b'_j]$ for some $b'_j \in \bools$ and if $\ket{\psi'} = \ket\psi_{i\rightarrow b'_j}$.
We define $ \mpred_{i \rightarrow j}(\varphi')$ as the set of~$i\rightarrow j$-predecessors of $\varphi'$.
\end{definition}

We now define the effect of measuring the $i$-th qubit into the~$j$-th register.
Intuitively, the probability of obtaining a hybrid state $(\ket\psi, \beta)$ with $\beta = b_1 \cdots b_m$ from the potential predecessor state $\varphi' = (\ket{\psi'}, \beta')$ by measuring the $i$-th qubit is the product of the probability $\rho(\varphi')$ of having obtained~$\varphi'$ beforehand and of the probability of measuring the $i$-th qubit in that state as~$b_j$, which we write as $\Prob_{\ket{\psi'}}(i \rightarrow b_j)$.

\begin{definition}[$\concapp{\meas\ i \rightarrow j}$]
  We define
  \begin{multline*}
      \concapp{\meas\ i \rightarrow j}(\rho) \colon (\varphi) \mapsto \\
      \Sigma_{(\ket{\psi'}, \beta') \in \mpred_{i \rightarrow j}(\varphi)} (\rho(\ket{\psi'}, \beta') \cdot \Prob_{\ket{\psi'}}(i \rightarrow b_j)) \enspace ,
  \end{multline*}
  where $ \beta = b_1 \cdots b_m$.
\end{definition}

The definition of~$\concapp{\meas\ i \rightarrow j}(\rho)$ contains a sum over all potential $i \rightarrow j$ predecessors of the given hybrid state.
As $\mpred_{i \rightarrow j}(\varphi)$ is, in general, infinite, it is not trivial to compute this sum.
We observe, however, that the summand reduces to~$0$ if~$\rho(\ket{\psi'}, \beta') = 0$.
Hence, as long as the machine state~$\rho$ assigns non-zero probabilities to only finitely many hybrid states, the machine state~$\rho'$ is straightforwardly computable.

We now proceed to define the effect of resetting the~$i$-th qubit to~$0$.
This can be implemented as a derived operation that is implemented by measuring the qubit and, if it is measured as~$1$, inverting it using an~$X$-gate.
This implementation, however, requires the measured value to be stored in some register although it is only used for the reset operation.
Hence, we define the reset-operation natively. %

We first define the effect of setting an individual qubit to a specific value.
While this operation is not possible in real-life quantum computers, it allows us a more straightforward formalization of the reset-operation later on.

\begin{definition}[Setting qubits]
  Let~$\ket\psi$ be a quantum state, let $i \in [n]$, and let~$b \in \bools$.
We write $\ket\psi[i\rightarrow b]$ to denote the quantum state~$\norm{f}$ with
\[
  f\colon (q) \mapsto \\ \begin{cases}
    \ket\psi(q) + \ket\psi(q[i \rightarrow 1-b]) & \text{if~$q_i = b$} \\
    0 & \text{otherwise,}
  \end{cases} \]
  where $q = q_1 \cdots q_n$.
\end{definition}

Using this operation allows us to define the $i{\downarrow}_b$-predecessors of a quantum state.
Intuitively, a quantum state~$\ket\psi$ is a $i{\downarrow}_0$-predecessor of a quantum state~$\ket{\psi'}$ if~$\ket{\psi'}$ results from~$\ket\psi$ by measuring the $i$-th qubit as~$0$.
Analogously, a quantum state~$\ket\psi$ is a $i{\downarrow}_1$-predecessor of a quantum state~$\ket{\psi'}$ if~$\ket{\psi'}$ results from~$\ket\psi$ by measuring the $i$-th qubit as~$1$ and that qubit is subsequently set to~$0$.

\begin{definition}[$i\downarrow_b$-predecessors]
Let $\varphi = (\ket\psi, \beta)$ with $\beta = b_1 \cdots b_m$ and $\varphi' = (\ket{\psi'}, \beta')$ with $\beta'= b'_1 \cdots b'_m$ be hybrid states.
We say that $\varphi$ is a \emph{$i{\downarrow}_0$-predecessor} of $\varphi'$ if $\beta' = \beta$ and if $\ket{\psi'} = \ket\psi_{i \rightarrow 0}$.
We say that $\varphi$ is a \emph{$i{\downarrow}_1$-predecessor} of $\varphi'$ if $\beta' = \beta$ and if $\ket{\psi'} = \ket{\psi}_{i \rightarrow 1}[i\rightarrow 0]$.
For $b \in \bools$ we define $\rpred_{i\downarrow_b}(\varphi')$ as the set of~$i\downarrow_b$-predecessors of $\varphi'$.
\end{definition}

This definition now allows us to define the effect of the reset operation on a machine state analogously to the application of a measurement operation.

\begin{definition}[$\concapp{\reset\ i}$]
  We define
  \begin{multline*}
    \concapp{\reset\ i}(\rho) \colon \varphi \mapsto \\
      \Sigma_{b \in \bools, (\ket{\psi'}, \beta') \in \rpred_{i\downarrow_b} (\varphi)} ( \rho(\ket{\psi'}, \beta') \cdot \Prob_{\ket{\psi'}}(i \rightarrow b) ) \enspace .
  \end{multline*}
\end{definition}

The above definitions provide us with a semantics of quantum circuits.
In the next section we discuss how to obtain information about the machine state without the overhead of concrete execution.

\subsection{Abstract Execution}
\label{sec:constant-propagation:abstract}

\newcommand{\ampl}{\text{Ampl}}
\newcommand{\absapp}[1]{[\![ #1 ]\!]^\#}

In the previous section we have formalized the state of a quantum computer and its evolution during the execution of a quantum circuit.
Each quantum machine state may assign non-zero probabilities to arbitrarily many hybrid states, each of which may comprise a sequence of $m$~registers and one quantum state containing up to $2^n$~many non-zero amplitudes.
Hence, the size of any individual quantum machine state is theoretically not bounded from above. %
To obtain constant propagation information efficiently, we impose a fixed upper limit on the size of quantum machine states in this section.

\begin{definition}[Representable states]
We call the set~$\ampl(\ket\psi) = \set{ \ket{\psi}(q) \mid q \in \bools^n, \ket{\psi}(q) \neq 0}$ the \emph{(non-zero) amplitudes} of~$\ket{\psi}$.
For~$N \in \nats$ we say that~$\ket\psi$ is \emph{$N$-representable} if~$\ket\psi$ has at most~$N$ non-zero amplitudes.
We say that the hybrid state~$(\ket\psi, \beta)$ is \emph{$N$-representable} if~$\ket\psi$ is $N$-representable.
We say that the machine state~$\rho$ is \emph{$N, M$-representable} if $\rho$ assigns a non-zero probability to at most~$M$ hybrid states.
\end{definition}

Our abstract domain consists mainly of~$N,M$-representable machine states together with the state~$\top$ that represents all non-$N,M$-representable states.

\begin{definition}[Abstract states]
We define the set of \emph{abstract hybrid states} as $\set{\top} \cup \set{\varphi \in \Phi \mid \Phi \text{ is $N$-representable}}$.
We define the set of \emph{abstract machine states} $\set{\top} \cup \set{\rho\in \mathcal P \mid \rho \text{ is $N, M$-representable}}$.
\end{definition}

Having defined the concrete application of operations via $\concapp{\cdot}$ we now proceed to define the abstract application.

\begin{definition}[Abstract execution]
  Let~$x \in \set{\app\ U, \capp\ U\ i_1\dots i_k, \meas\ i \rightarrow j, \reset\ i}$.
  We define the abstract execution function~$\absapp{x}$ as follows:
  \[
    \absapp{x}(\rho) = \begin{cases}
      \top &\text{if } \rho = \top \\
      \top &\text{if } \concapp{x}(\rho) \text{ is not $N,M$-representable} \\
      \concapp{x}(\rho) &\text{otherwise}
    \end{cases}
  \]
\end{definition}

\subsection{Efficient Machine State Storage via Hybrid Union Tables}

In the last section we restricted the size of a machine state to not exceed a pre-defined threshold to counter the inherent exponential growth of quantum states~\cite{feynman1982simulating}.
In addition to this artificial upper bound, we aim to use the available space for keeping machine states as efficiently as possible.
Chen and Stade~\cite{chen2023quantum} introduced union tables to efficiently store quantum states, which we now extend to efficiently store machine states.

The key insight is that one can store the amplitudes of non-entangled sets of qubits separately.
Consider, e.g., a simple quantum circuit with three qubits, each of which has a Hadamard-gate applied to it.
This yields the quantum state~$\ket\psi$ with $\ket\psi(q) = \nicefrac{1}{\sqrt{8}}$ for each~$q \in \bools^3$.
Storing~$\ket\psi$ explicitly would require storing eight amplitudes, one for each combination of the results of measuring the three qubits individually.
As the three qubits are not entangled, however, it suffices to consider each as an individual quantum system.
This allows us to store the amplitudes of each qubit individually, resulting in six stored amplitudes, two for each qubit.

\begin{figure}
	
	\pdftooltip{
  	\begin{quantikz}[wire types={q,q,q,q,q,c,c},row sep={.5cm,between origins}]
		\lstick{$q_1$} & \gate{H} & \meter{} \wire[d][5]{c} & \qw & \qw & \qw & \qw & \qw & \qw\\
		\lstick{$q_2$} & \qw & \qw & \qw & \qw & \qw & \qw & \targ{} & \qw\\
		\lstick{$q_3$} & \gate{H} & \qw & \qw & \meter{} \wire[d][4]{c} & \gate{X} & \gate{H} & \ctrl{-1} & \qw\\
        \lstick{$q_4$} & \qw & \qw & \gate{X} & \qw & \qw & \qw & \qw & \qw\\
		\lstick{$q_5$} & \gate{H} & \qw & \qw & \qw & \qw & \gate{Z} & \qw & \qw\\
        \lstick{$r_1$} & & \control{} & \control{} \wire[u][2]{c} & & & & &\\
        \lstick{$r_2$} & & & & \control{} & \control{} \wire[u][4]{c} & & &
	\end{quantikz}
	}{A quantum circuit with five qubits and two bits. The applied gates are (in order): A Hadamard gate applied to the first, a Hadamard gate applied to the third qubit, and a Hadamard gate applied to the fifth qubit. The first qubit is measured, the result is written into the first bit. A NOT controlled classically by the first bit is applied to the fourth qubit. The third qubit is measured, the result is written into the second bit. A NOT controlled classically by the second bit is applied to the third qubit. A Hadamard gate applied to the third qubit, and a Pauli-Z gate applied to the fifth qubit. A CNOT gate with the third qubit as control and second as target.}
	\caption{%
    An example of a hybrid quantum circuit. %
    We show the resulting union table in \Cref{fig:qcpExampleTable}. %
  	}
	\label{fig:qcpExampleCircuit}
\end{figure}
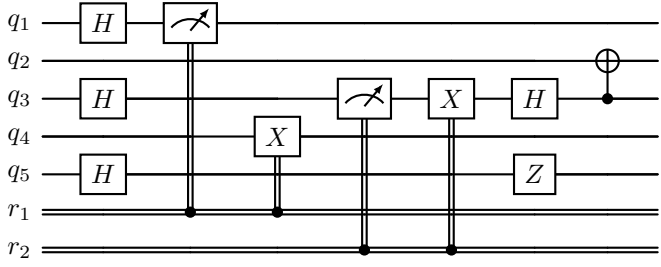

\begin{figure}
	\pdftooltip{
	\parbox{\columnwidth} {%
	\centering
  	\begin{tikzpicture}[thick,yscale=.9]

  \node[draw,anchor=north,minimum height=2.75cm,minimum width=.5cm,label=above:Qubits] (qubits-frame) at (0,0) {};

  \node (q1) at ($(qubits-frame.north) ! .1 ! (qubits-frame.south)$) {$q_1$};
  \node (q2) at ($(qubits-frame.north) ! .3 ! (qubits-frame.south)$) {$q_2$};
  \node (q3) at ($(qubits-frame.north) ! .5 ! (qubits-frame.south)$) {$q_3$};
  \node (q4) at ($(qubits-frame.north) ! .7 ! (qubits-frame.south)$) {$q_4$};
  \node (q5) at ($(qubits-frame.north) ! .9 ! (qubits-frame.south)$) {$q_5$};

  \node[draw,minimum width=6cm,minimum height=1cm,anchor=north west] (union-table-1) at ($(qubits-frame.north east) + (.5cm,0)$) {};
    \node[] at ($(union-table-1) + (-0.9,.25)$) {$\ket{00} \rightarrow 1$};
    \node[] at ($(union-table-1) + (1.67,.25)$) {$0$};
    \node[] at ($(union-table-1) + (2.55,.25)$) {$0.5$};

    \node[] at ($(union-table-1) + (-0.9,-.25)$) {$\ket{11} \rightarrow 1$};
    \node[] at ($(union-table-1) + (1.67,-.25)$) {$1$};
    \node[] at ($(union-table-1) + (2.55,-.25)$) {$0.5$};

  \draw[thick,dotted] ($(union-table-1.north west) ! .7 ! (union-table-1.north east)$) -- ($(union-table-1.south west) ! .7 ! (union-table-1.south east)$);
  \draw[thick,dotted] ($(union-table-1.north west) ! .85 ! (union-table-1.north east)$) -- ($(union-table-1.south west) ! .85 ! (union-table-1.south east)$);
  \draw[thick,dotted] (union-table-1.west) -- (union-table-1.east);

  \node[draw,minimum width=6cm,minimum height=1cm,anchor=north] (union-table-2) at ($(union-table-1.south) + (0,-.1cm)$) {};
    \node[] at ($(union-table-2) + (-0.9,.25)$) {$\ket{00} \rightarrow \nicefrac{1}{\sqrt{2}};\ket{11} \rightarrow \nicefrac{1}{\sqrt{2}}$};
    \node[] at ($(union-table-2) + (1.67,.25)$) {$0$};
    \node[] at ($(union-table-2) + (2.55,.25)$) {$0.5$};

    \node[] at ($(union-table-2) + (-0.9,-.25)$) {$\ket{00} \rightarrow \nicefrac{1}{\sqrt{2}};\ket{11} \rightarrow \nicefrac{1}{\sqrt{2}}$};
    \node[] at ($(union-table-2) + (1.67,-.25)$) {$1$};
    \node[] at ($(union-table-2) + (2.55,-.25)$) {$0.5$};

  \draw[thick,dotted] ($(union-table-2.north west) ! .7 ! (union-table-2.north east)$) -- ($(union-table-2.south west) ! .7 ! (union-table-2.south east)$);
  \draw[thick,dotted] ($(union-table-2.north west) ! .85 ! (union-table-2.north east)$) -- ($(union-table-2.south west) ! .85 ! (union-table-2.south east)$);
  \draw[thick,dotted] (union-table-2.west) -- (union-table-2.east);

  \node[draw,minimum width=6cm,minimum height=.5cm,anchor=north] (union-table-3) at ($(union-table-2.south) + (0,-.1cm)$) {};
    \node[] at ($(union-table-3) + (-0.9,0)$) {$\ket{0} \rightarrow \nicefrac{1}{\sqrt{2}};\ket{1} \rightarrow \nicefrac{-1}{\sqrt{2}}$};
    \node[] at ($(union-table-3) + (2.55,0)$) {$1.0$};

  \draw[thick,dotted] ($(union-table-3.north west) ! .7 ! (union-table-3.north east)$) -- ($(union-table-3.south west) ! .7 ! (union-table-3.south east)$);
  \draw[thick,dotted] ($(union-table-3.north west) ! .85 ! (union-table-3.north east)$) -- ($(union-table-3.south west) ! .85 ! (union-table-3.south east)$);

  \node[draw,minimum width=.5cm,minimum height=1cm,anchor=north west,label=above:Registers] (bits-frame) at ($(union-table-1.north east) + (.5cm,0)$) {};

  \node (b1) at ($(bits-frame.north) ! .25 ! (bits-frame.south)$) {$r_1$};
  \node (b2) at ($(bits-frame.north) ! .75 ! (bits-frame.south)$) {$r_2$};

  \path[draw,rounded corners=2pt,-stealth] (q1 -| qubits-frame.east) -- ++(.15cm,0) |- (union-table-1.west);
  \path[draw,rounded corners=2pt,-stealth] (q4 -| qubits-frame.east) -- ++(.15cm,0) |- (union-table-1.west);

  \path[draw,rounded corners=2pt,-stealth] (q2 -| qubits-frame.east) -- ++(.3cm,0) |- (union-table-2.west);
  \path[draw,rounded corners=2pt,-stealth] (q3 -| qubits-frame.east) -- ++(.3cm,0) |- (union-table-2.west);

  \path[draw,rounded corners=2pt,-stealth] (q5 -| qubits-frame.east) -- (union-table-3.west);

  \path[draw,rounded corners=2pt,-stealth] (b1 -| bits-frame.west) -- ++(-.15cm,0) |- (union-table-1.east);
  \path[draw,rounded corners=2pt,-stealth] (b2 -| bits-frame.west) -- ++(-.3cm,0) |- (union-table-2.east);

  \path[draw,decorate,decoration={brace,amplitude=4pt}]
    (union-table-1.north west)
    -- node[anchor=south,yshift=2pt] {Quantum States}
    ($(union-table-1.north west) ! .7 ! (union-table-1.north east)$);
  \path[draw,decorate,decoration={brace,amplitude=4pt}]
    ($(union-table-1.north west) ! .7 ! (union-table-1.north east)$)
    -- node[anchor=south,align=center,yshift=2pt] {Bit Se-\\quence}
    ($(union-table-1.north west) ! .85 ! (union-table-1.north east)$);
  \path[draw,decorate,decoration={brace,amplitude=4pt}]
    ($(union-table-1.north west) ! .85 ! (union-table-1.north east)$)
    -- node[anchor=south,align=center,yshift=2pt] {$\rho$}
    (union-table-1.north east);

  \path[draw,decorate,decoration={brace,amplitude=4pt,mirror}]
    (union-table-3.south west)
    -- node[anchor=north,yshift=-2pt] {Hybrid States}
    ($(union-table-3.south west) ! .85 ! (union-table-3.south east)$);

\end{tikzpicture}%
  	}%
  	}{A depiction of an example union table. Qubit indices and register indices refer to entries in the table.
  	Qubit index 1 and 4 as well as register index 1 refer to the first table entry. The entry contains of two hybrid states, both with a probability rho of 0.5. The first hybrid state contains of the quantum state |00> -> 1 and the bit sequence 0. The second hybrid state contains of the quantum state |11> -> 1 and the bit sequence 1.
	Qubit index 2 and 3 as well as register index 2 refer to the second table entry. The entry contains of two hybrid states, both with a probability rho of 0.5. The first hybrid state contains of the quantum state |00> -> 1/sqrt(2); |11> -> 1/sqrt(2) and the bit sequence 0. The second hybrid state contains of the quantum state |00> -> 1/sqrt(2); |11> -> 1/sqrt(2) and the bit sequence 1.
  	Qubit index 5 refers to the third table entry. The entry contains of one hybrid state, with a probability rho of 1. The hybrid state contains of the quantum state |0> -> 1/sqrt(2); |1> -> -1/sqrt(2).}
	\caption{The union table that results from the circuit in \Cref{fig:qcpExampleCircuit}. %
    Adapted from \cite[Figure~3]{chen2023quantum}.}%
	\label{fig:qcpExampleTable}
\end{figure}

We extend this observation to the hybrid setting by considering a register~$i$ ``entangled'' with a qubit~$j$ if we
\begin{enumerate*}[label=(\alph*)]
  \item measure the value of qubit~$j$ into register~$i$ or if we
  \item use register~$i$ to guard the application of a gate to qubit~$j$.
\end{enumerate*}
This allows us to store a machine state in a hybrid union table.

For the sake of brevity we omit a formal definition and instead refer to the example circuit shown in \Cref{fig:qcpExampleCircuit} and its resulting union table shown in \Cref{fig:qcpExampleTable}.
That union table contains three entries which characterize qubits~$q_1$ and~$q_4$ together with register~$r_1$, qubits~$q_2$ and~$q_3$ together with register~$r_2$, and qubit~$q_5$ without any registers, respectively.
If later on qubit~$q_1$ and~$q_3$ were to be entangled, the first and second entry of the union table would be merged as described by Chen and Stade~\cite{chen2023quantum}.

\section{Transformation Routines\\*via Constant Propagation}\label{sect:transformations-qcp}

Using the constant propagation of the last section, we can apply transformation routines to the quantum circuit, which do not change the semantics.
In this section, we will introduce these transformations.
For the sake of brevity, we introduce notation for the case that a quantum state~$\ket\psi$ assigns the value~$b$ to the~$i$-th qubit:
We define $\ket\psi_i = b \Leftrightarrow \forall b_1\cdots b_n \in \bools^n.\, b_i = 1-b \Rightarrow \ket\psi(b_1\cdots b_n) = 0$.

\paragraph*{General Control Reduction}

Chen and Stade~\cite{chen2023quantum} argued that quantum constant propagation can be used to reduce the number of controlled gates as follows.
If we know that a controlling qubit is always~$0$ ($1$), we can remove the controlled gate (the control).
By including registers into our analysis, we extend this principle to classical controls of an instruction.
If we know that a controlling register will always be $0$ ($1$), we can remove the instruction (the control).

\paragraph*{Unsatisfiable Controls}

As Chen and Stade pointed out, quantum constant propagation can also remove gates with multiple controls, if the combination of controls is not satisfiable~\cite{chen2023quantum}.
We extend this principle to classical controls.

\paragraph*{Quantum Control Reduction}

A key goal of hybrid quantum constant propagation is to replace quantum resources by classical ones.
One possible avenue for this is to replace qubit-controlled gates by register-controlled ones.
This is done by checking whether there exist~$i \in [n]$, $j \in [m]$ and~$b,b' \in \bools$ such that $\ket{\psi}_i = b \Leftrightarrow r_j = b'$.
In either case, a gate controlled by qubit~$q_i$ can be replaced by one controlled (either positively or negatively) by register~$r_j$.

This generalizes the measurement lifting depicted in \Cref{fig:dynamicCircuitExample}.
It allows for routines like the one of \Cref{fig:hcpExample}, without depending on pattern-matching certain circuit structures.

\paragraph*{Implied Qubits/Registers}

If a gate is controlled by multiple registers or qubits, we can remove some of the controls by checking for implications.

If there exist~$i,j \in [n]$ and~$b, b' \in \bools$ such that $q_i$ and $q_j$ control the same gate and such that $\ket{\psi}_i = b \Rightarrow \ket{\psi}_j = b'$, we can remove either $q_i$ or~$q_j$ from the controls.
Analogously, if for $i \in [n]$ and $j \in [m]$ we have that $q_i$ and $r_j$ control the same gate and $r_j = b \Rightarrow \ket\psi_i = b'$, we can remove $q_i$ from the controls.

\paragraph*{Phase Gate Reduction}

Phase gates are matrices that only have non-zero entries on their diagonal.
If the phase gate applies the same phase to all non-zero amplitudes of the quantum state, the phase gate only applies a global phase and can be deleted.
Instead, the global phase is added to the circuit, as long as it is not $1$.

\section{Experimental Evaluation}\label{sect:experimental-evaluation}

The proposed implementation passes have been implemented on top of the MQT Compiler Collection~\cite{burgholzer2026mqt} as part of the MQT Core project~\cite{burgholzer2025MQTCore,mqt_core_fork}, which is part of the MQT~\cite{wille2024mqt}.
The implementation uses the \ac{mlir} framework~\cite{lattner2021mlir,hopf2026integrating}.
We evaluated measurement lifting~\cite{rovara2025qubit}, measurement lifting with Hadamard lifting (see \Cref{sec:hadamard-lifting}), constant propagation (see \Cref{sec:constant-propagation}), and the combination of constant propagation and measurement lifting. 
We also applied quantum constant propagation as proposed by Chen and Stade~\cite{chen2023quantum}.
In this case, we did not implement the tool in MQT Core, but used the tool provided by them.

All passes were applied until a fixed point. 
When multiple passes were used they were applied alternately until convergence.
Constant propagation used 16,4-representable machine states, i.e., we track at most~$4$ hybrid states, each of which comprises at most~$16$ non-zero amplitudes.
Analogously, we allowed at maximum 16 nonzero amplitudes for quantum constant propagation.

We ran the optimization routines on the benchmark circuits of MQT Bench~\cite{quetschlich2023mqtbench}.
To use the circuits, we removed all \texttt{barrier} instructions.
This is because they are mostly used for visual separation of different circuit parts in MQT~Bench.
Additionally, they would---as is their functionality---prevent us from applying the optimization techniques effectively and showing whether the techniques work in principle or not.

We ended up with $3\thinspace 797$~circuits to benchmark.
The circuits implemented the following algorithms:
\begin{description}
  \item[AE] Amplitude estimation ($2$~to $19$~qubits)~\cite{brassard2000quantum},
  \item[BV] Bernstein-Vazirani ($2$~to $200$~qubits)~\cite{bernstein1993quantum},
  \item[CDKM] CDKM ripple-carry adder circuit ($2$ to $200$~qubits, even numbers only)~\cite{vedral1996quantum,cuccaro2004new},
  \item[DJ] Deutsch Josza ($2$ to $200$~qubits)~\cite{deutsch1992rapid},
  \item[FA] quantum full adders ($4$ to $200$~qubits, even numbers only),
  \item[GHZ] the creation of a GHZ state ($2$~to $200$~qubits)~\cite{greenberger1989going},
  \item[GS] the representation of a graph state ($3$~to $200$~qubits), 
  \item[Grover] Grover's algorithm ($2$ to $20$~qubits)~\cite{grover1996fast},
  \item[HA] quantum half adders ($3$ to $199$~qubits, odd numbers only),
  \item[HHL] Harrow–Hassidim–Lloyd ($3$ to $200$~qubits)~\cite{harrow2009quantum},
  \item[HRS] HRS Cumulative Multiplier ($5, 9, 13,\dots$ $197$~qubits)~\cite{haner2018optimizing},
  \item[MA] modular adder circuits ($2$ to $200$~qubits, even numbers only),
  \item[Multiplier] quantum multipliers ($4, 8, 12,\dots 108$~qubits),
  \item[QAOA] quantum approximate optimization algorithm ($2$ to $200$~qubits)~\cite{farhi2014quantum},
  \item[QFT] quantum fourier transform ($2$ to $200$~qubits)~\cite{draper2000addition},
  \item[QFTa] Draper QFT Adder ($2$ to $200$~qubits, even numbers only)~\cite{draper2000addition},
  \item[QFTe] QFT with a previous entangling routine for qubits ($2$ to $20$ and $100$~qubits),
  \item[QFTr] RG QFT multiplier ($4, 8, 12,\dots 92$~qubits)~\cite{ruiz2017quantum},
  \item[QNN] quantum neural networks ($2$~to $200$~qubits)~\cite{qnn_qiskit},
  \item[QPEe/QPEi] quantum phase estimation with an exact and inexact representation of the phase ($2$~to $200$~qubits each)~\cite{kitaev1995quantum},
  \item[QUARK] a cardinality and a copula circuit from the modeling application in the QUARK framework ($2$ to $200$~qubits, copula circuits have an even number of qubits only)~\cite{finvzgar2022quark,quark_github},
  \item[QWalk] quantum walk ($3$~qubits)~\cite{kempe2003quantum},
  \item[Shor] Shor's algorithm ($18$ and $42$~qubits)~\cite{shor1994algorithms},
  \item[VBE] VBE ripple carry adder ($4, 7, 10,\dots 199$~qubits)~\cite{vedral1996quantum},
  \item[VQE] variational quantum eigensolver with SU(2) ansatz~\cite{qiskit_su_2}, real amplitudes ansatz~\cite{qiskit_real_amplitudes}, and two local ansatz~\cite{qiskit_two_local}  ($2$~to $200$~qubits each)~\cite{peruzzo2014variational}
  \item[WState] the creation of a W-State ($2$~to $200$~qubits)~\cite{dur2000three}.
\end{description}

\begin{figure*}
	\pdftooltip{
	\parbox{\textwidth} {%
	\centering
	\includegraphics[width=\textwidth]{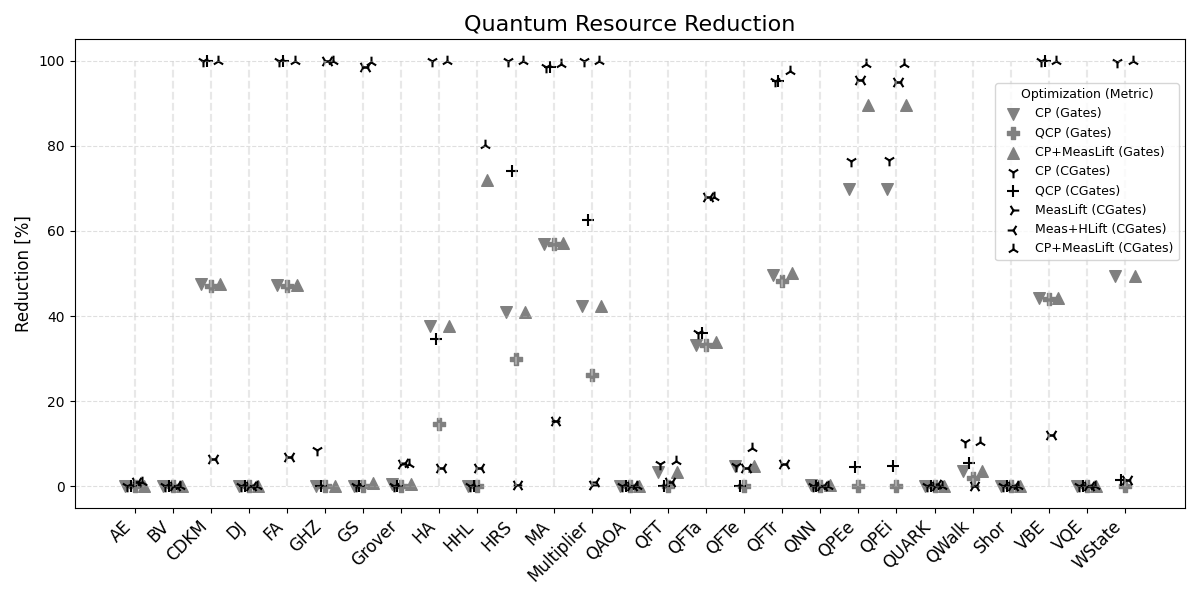}
	}%
	}{A graph titled “Quantum Resource Reduction”. It has multiple algorithms on the x-axis the reduction in percentage on the y-axis. The reductions are shown for CP (Gates), QCP (Gates), CP+MeasLift (Gates), CP (CGates), QCP (CGates), MeasLift (CGates), Meas+HLift (CGates), and CP+MeasLift (CGates). The values for each algorithm are, in the previous given order of optimization metrics:
	AE	0.00\%	0.00\%	0.00\%	0.00\%	0.00\%	1.00\%	1.00\%	1.00\%; 
	BV	0.00\%	0.00\%	0.00\%	0.00\%	0.00\%	0.00\%	0.00\%	0.00\%; 
	CDKM	47.44\%	47.14\%	47.44\%	100.00\%	100.00\%	6.55\%	6.55\%	100.00\%; 
	DJ	0.00\%	0.00\%	0.00\%	0.00\%	0.00\%	0.00\%	0.00\%	0.00\%; 
	FA	47.40\%	47.12\%	47.40\%	100.00\%	100.00\%	6.81\%	6.81\%	100.00\%; 
	GHZ	0.00\%	0.00\%	0.00\%	8.58\%	0.00\%	100.00\%	100.00\%	100.00\%; 
	GS	0.00\%	0.00\%	0.83\%	0.00\%	0.00\%	98.53\%	98.53\%	99.75\%; 
	Grover	0.66\%	0.00\%	0.66\%	0.00\%	0.00\%	5.33\%	5.43\%	5.33\%; 
	HA	37.57\%	14.76\%	37.57\%	100.00\%	34.58\%	4.43\%	4.43\%	100.00\%; 
	HHL	0.00\%	0.00\%	72.03\%	0.00\%	0.00\%	4.42\%	4.42\%	80.14\%; 
	HRS	40.94\%	29.99\%	40.94\%	100.00\%	74.19\%	0.24\%	0.24\%	100.00\%; 
	MA	56.88\%	56.87\%	57.21\%	98.55\%	98.54\%	15.41\%	15.41\%	99.28\%; 
	Multiplier	42.27\%	26.22\%	42.27\%	100.00\%	62.56\%	0.36\%	1.14\%	100.00\%; 
	QAOA	0.00\%	0.00\%	0.00\%	0.00\%	0.00\%	0.00\%	0.00\%	0.00\%; 
	QFT	3.48\%	0.00\%	3.48\%	5.20\%	0.00\%	1.15\%	1.15\%	5.93\%; 
	QFTa	33.33\%	33.33\%	33.81\%	35.97\%	35.97\%	67.98\%	67.98\%	67.98\%; 
	QFTe	4.76\%	0.00\%	4.76\%	4.90\%	0.00\%	4.24\%	4.24\%	9.14\%; 
	QFTr	49.70\%	48.24\%	50.01\%	95.31\%	95.31\%	5.34\%	5.34\%	97.66\%; 
	QNN	0.31\%	0.00\%	0.31\%	0.03\%	0.00\%	0.00\%	0.00\%	0.03\%; 
	QPEe	69.81\%	0.00\%	89.52\%	76.54\%	4.55\%	95.40\%	95.40\%	99.27\%; 
	QPEi	69.94\%	0.00\%	89.68\%	76.58\%	4.90\%	95.08\%	95.08\%	99.23\%; 
	QUARK	0.00\%	0.00\%	0.00\%	0.00\%	0.00\%	0.00\%	0.47\%	0.00\%; 
	QWalk	3.57\%	1.99\%	3.57\%	10.36\%	5.48\%	0.04\%	0.08\%	10.39\%; 
	Shor	0.00\%	0.00\%	0.00\%	0.04\%	0.04\%	0.15\%	0.15\%	0.18\%; 
	VBE	44.15\%	43.91\%	44.15\%	100.00\%	100.00\%	12.07\%	12.07\%	100.00\%; 
	VQE	0.00\%	0.00\%	0.00\%	0.03\%	0.00\%	0.00\%	0.00\%	0.03\%; 
	WState	49.45\%	0.00\%	49.45\%	99.75\%	1.49\%	1.48\%	1.48\%	100.00\%; 
	}
	\caption{The amount of quantum resources reduced on average in circuits implementing different algorithms by quantum-classical constant propagation (CP), quantum constant propagation (QCP), measurement lifting (MeasLift), measurement lifting with Hadamard lifting (Meas+HLift), and constant propagation with measurement lifting (QCP+MeasLift). ``Gates'' is the reduction of whole quantum gates, while ``CGates'' is the reduction of controlled gates. The latter is both the complete removal of the gates as well as turning the quantum controlled gates into non-quantum controlled gates.}\%
	\label{fig:gateReduction}
\end{figure*}

\begin{figure*}
	\pdftooltip{
		\parbox{\textwidth} {%
			\centering
			\includegraphics[width=\textwidth]{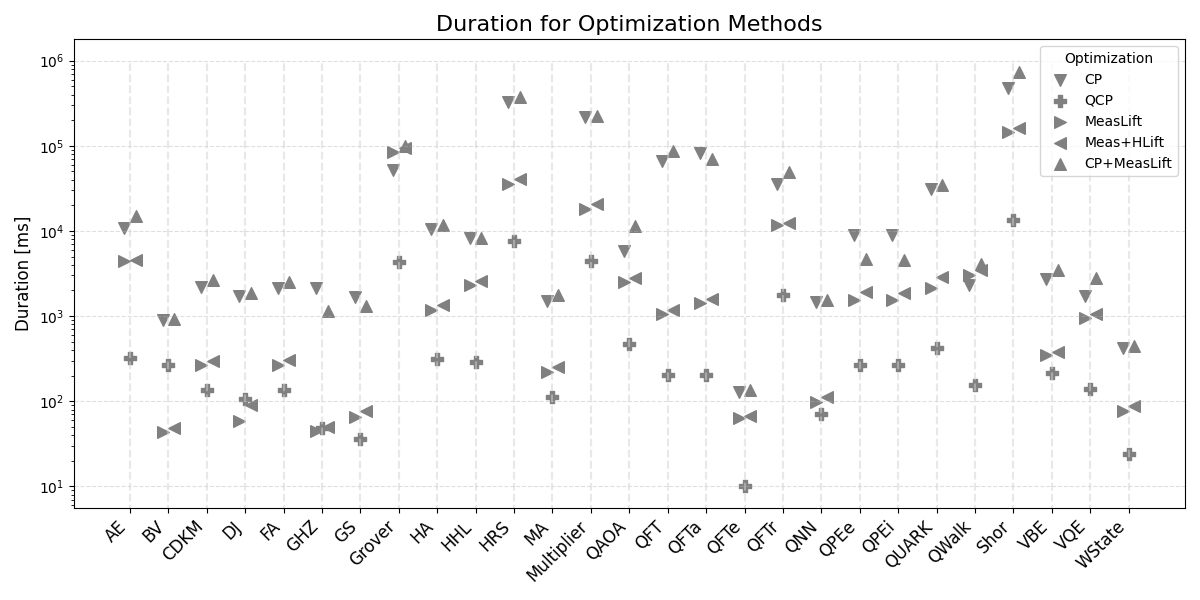}
		}%
	}{A graph titled “Duration for Optimization Methods”. It has multiple algorithms on the x-axis the duration in ms on the y-axis. The durations are shown for CP, QCP, CP+MeasLift, Meas+HLift, and CP+MeasLift. The values for each algorithm are, in the previous given order of optimization metrics:
	AE	10882	319	4498	4558	15115; 
	BV	901	265	44	49	922; 
	CDKM	2173	134	267	301	2672; 
	DJ	1702	106	59	90	1883; 
	FA	2156	135	266	302	2524; 
	GHZ	2157	49	45	50	1142; 
	GS	1661	36	66	76	1302; 
	Grover	51752	4289	84740	94684	99789; 
	HA	10593	315	1175	1338	11844; 
	HHL	8201	290	2316	2593	8209; 
	HRS	327811	7549	36043	40595	370306; 
	MA	1488	111	222	252	1753; 
	Multiplier	215606	4395	18135	20717	222835; 
	QAOA	5863	468	2548	2810	11430; 
	QFT	65746	201	1051	1172	87963; 
	QFTa	82999	203	1436	1600	69074; 
	QFTe	127	10	63	68	136; 
	QFTr	36071	1785	11705	12392	49287; 
	QNN	1478	71	99	111	1527; 
	QPEe	8856	269	1560	1911	4683; 
	QPEi	8858	267	1551	1887	4521; 
	QUARK	30819	420	2131	2854	34785; 
	QWalk	2289	154	3001	3442	4132; 
	Shor	482204	13475	145102	162500	739590; 
	VBE	2725	215	345	379	3443; 
	VQE	1735	140	939	1059	2832; 
	WState	419	24	76	88	444; 
	}
	\caption{The duration in ms needed to apply the optimization techniques quantum-classical constant propagation (CP), quantum constant propagation (QCP), measurement lifting (MeasLift), measurement lifting with Hadamard lifting (Meas+HLift), and constant propagation with measurement lifting (QCP+MeasLift). It is noteworthy that the usage of constant propagation makes the compilation much longer. Measurement lifting is quicker than measurement lifting with Hadamard lifting. In some cases, constant propagation and measurement lifting is quicker than constant propagation alone. The comparison with quantum constant propagation should be taken with care, as this routine has been executed on another framework without casting to MLIR.}%
	\label{fig:duration}
\end{figure*}

We show detailed results and the relative reduction of quantum resources achieved in \Cref{fig:gateReduction}.
The effectiveness of the optimization techniques depends heavily on the targeted circuits.
In some algorithms, quantum resources decrease drastically, while in others, no change occurred.

The results show that there are algorithms which benefit from quantum-classical constant propagation, namely HA, HRS, quantum multiplier, QFT, QFTr, both QPEs, and circuits for the creation of the WState.
In all cases, the hybrid constant propagation leads to a larger amount of reduced quantum resources than the other optimization techniques.

For other algorithms, we could not find an advantage of using a hybrid approach.
Here, no difference could be detected between a hybrid and a full-quantum constant propagation.
This applies to CDKM, FA, MA, QFTa, and VBE.

Measurement lifting does not reduce the total amount of gates, but can reduce the amount of controlled quantum gates.
In some cases, this leads to a larger reduction of controlled gates than constant propagation can achieve.
We observe this for GHZ, GS, Grover, QFTa, and both QPEs.

Applying Hadamard lifting to circuits on top of measurement lifting did not or only minimally improve the results.
This could mean that the patterns targeted by Hadamard lifting do not appear in today's quantum algorithms.

The combination of measurement lifting and constant propagation can remove more quantum resources than using one algorithm on its own.
This is the case for HHL, QFTe, and both QPEs.
We assume the reason for this to be that constant propagation can find removable quantum resources which pattern matching approaches do not detect.
It reduces quantum resources at the start of the circuits, and becomes worse as the circuit progresses.
On the other hand, measurement lifting is especially powerful at the end, where all qubits are measured.
Therefore, there can be circuits where both algorithms find distinct quantum resources to remove, which makes their combination especially powerful.

A special case in this regard is the HHL algorithm.
Here, all individual optimization techniques yield almost no quantum resource reduction.
In contrast, the combination of constant propagation and measurement lifting reduces quantum resources on average by over $70\%$.
This is because measurement lifting reorders gates and measurements, which does not reduce the quantum resources.
However, after reordering, constant propagation can be applied effectively.

For some algorithms, the application of constant propagation or measurement lifting results in a complete removal of controlled quantum gates.
This has different reasons.
One reason it that controls are zero if the qubits are $\ket{0}$ at the start of the circuit, as we assume for constant propagation.
The circuits that represent additions or multiplications (CDKM, FA, HA, HRS, VBE) get heavily reduced in the number of controls in this case.
In these cases, we need to keep in mind that the results are different if the circuits are sub-circuits of larger ones and do not start with all qubits in $\ket{0}$ state.

In GHZ, QPE and WState, the removal of the controlled gates roots in a mix of measurements that directly follow controls (removed by measurement lifting), controls that are determined by registers and can become classically controlled, and controlled phase gates that add only a global phase.

In general, our findings show that hybrid constant propagation is a promising addition to existing algorithms.
It can lead to the reduction of quantum resources when other optimization techniques fail, and enhances a compiler's ability to de-quantize a quantum circuit to make it more robust.

However, it needs to be pointed out that the compilation time of constant propagation is much larger than the one of measurement lifting.
This is depicted in \Cref{fig:duration}.

For all algorithms except Grover, constant propagation increases the compilation time greatly.
One way to deal with this could be to apply constant propagation to small and therefore quickly compiled versions of an algorithm.
If the results of this routine are promising (i.\,e. quantum resources are reduced), there are two options:
Constant propagation can be applied to a larger circuit and the longer compilation time is accepted, or the cases in which constant propagation can remove a gate are evaluated and applied to a larger circuit via pattern matching.
In the second case, one has to make sure not to remove non-superfluous quantum resources unwillingly.

One observation is that, for some circuits, the combination of measurement lifting and constant propagation is faster than the execution of constant propagation alone.
This could be explained by measurement lifting applying changes early that would be applied in later constant propagation iterations, causing a longer calculation time until a fixed point is reached.

\Cref{fig:duration} shows that quantum constant propagation usually runs faster than other evaluated optimization techniques.
This result has to be viewed with care, as we used a different tool for this routine.
That tool did not use \ac{mlir}, hence it can be less easily integrated in a compilation stack.

\section{Conclusion}\label{sect:conclusion}

We presented quantum-classical constant propagation as a strategy for reducing the quantum resources within a quantum circuit.
We were driven by the practical constraints of \ac{nisq} devices:
qubit operations, especially multi-qubit operations, are slow and error-prone.
Therefore, removing unnecessary quantum resources yields direct gains in fidelity and runtime.

We combined our techniques with measurement lifting~\cite{rovara2025qubit} and implemented them in MQT Core~\cite{burgholzer2025MQTCore,mqt_core_fork}.

We showed that quantum-classical constant propagation can effectively reduce quantum resources in cases where measurement lifting or quantum constant propagation cannot.
We also observed that constant propagation takes a long time and does not guarantee reduced quantum resources.

We therefore suggest that a reduction potential of constant propagation can be examined by applying the routine on small circuits.
We propose to apply constant propagation on a small test circuit of the algorithm under analysis.
If the quantum resources of this circuit can be reduced, it could be worthwhile to also apply constant propagation to larger circuits of the same algorithm.
Alternatively, one could analyze the changes introduced by constant propagation and find patterns to apply instead of running constant propagation on the complete circuit.
In this case, one needs to be careful not to unwittingly change the semantics.

We could also show that in some scenarios constant propagation and measurement lifting together are more effective than each routine on its own.
We could, however, not show that Hadamard lifting improves measurement lifting.

We conclude that quantum-classical constant propagation, measurement lifting, and their combination, can be useful tools to improve quantum circuits, especially in the \ac{nisq} era.
It can ease qubit mapping pressure, and lower calculation times as well as error rates on current hardware.
By selectively outsourcing work to classical control when it is safe to do so, we keep quantum resources focused where they deliver value.
This focused use of quantum hardware is a practical step toward reliable, scalable quantum advantage.

\section*{Acknowledgment}

We would like to thank Damian Rovara for the fruitful discussions that enriched this work, as well as for generously providing code templates.

Generative AI has been used in this work to create first code drafts (Microsoft Copilot) and review human-written text (GPT-4.1).

This work has been funded by DLR QCI via the CLIQUE project, the European Research Council (ERC) under the European Union’s Horizon 2020 research and innovation program grant agreement No. 101001318, and the Munich Quantum Valley (MQV K5 + K7), which is supported by the Bavarian state government with funds from the Hightech Agenda Bayern Plus.

\printbibliography

@STRING{natcomms = {Nature Communications} }

@STRING{sam 	= {IEEE Sensor Array and Multichannel Signal Processing Workshop (SAM)} }

@STRING{qce	= {IEEE International Conference on Quantum Computing and Engineering (QCE)} }

@STRING{qsw	= {IEEE International Conference on Quantum Software (QSW)} }

@STRING{sca_hpcasia = {Proceedings of the Supercomputing Asia and International Conference on High Performance Computing in Asia Pacific Region} }

@STRING{qic	= {Quantum Information \& Computation (QIC)} }

@STRING{tools	= {International Conference on Objects, Models, Components, Patterns (TOOLS)} }

@STRING{date	= {Design, Automation and Test in Europe (DATE)} }

@STRING{dac	= {Design Automation Conference (DAC)} }

@STRING{physreviewa   = {Physical Review A} }

@STRING{physreviewresearch   = {Physical Review Research} }

@STRING{joss = {Journal of Open Source Software} }

@STRING{aps = {APS Global Physics Summit} }

@STRING{qip	= {Quantum Information Processing} }

@STRING{is	= {IEEE Software} }

@STRING{gecco	= {Genetic and Evolutionary Computation Conference} }

@STRING{stoc	= {Symposium on Theory of Computing} }

@article{beauregard2002circuit,
	title={Circuit for {S}hor's algorithm using 2n+3 qubits},
	author={Beauregard, Stephane},
	year = {2003},
	issue_date = {March 2003},
	publisher = {Rinton Press, Incorporated},
	address = {Paramus, NJ},
	volume = {3},
	number = {2},
	journal = qic,
%	month = mar,
	pages = {175–185},
	numpages = {11}
}

@article{burgholzer2026mqt,
	title        = {The {{MQT Compiler Collection}}: {{A}} Blueprint for a Future-Proof Quantum-Classical Compilation Framework},
	author       = {Burgholzer, Lukas and Haag, Daniel and Stade, Yannick and Rovara, Damian and Hopf, Patrick and Wille, Robert},
	year         = {2026},
	booktitle    = date,
	eprint       = {2604.08674},
	eprinttype   = {arxiv},
}

@article{brassard2000quantum,
	title={Quantum {A}mplitude {A}mplification and {E}stimation},
	author={Brassard, Gilles and Hoyer, Peter and Mosca, Michele and Tapp, Alain},
	journal={arXiv preprint quant-ph/0005055},
	year={2000}
}

@article{briegel2009measurement,
  title={Measurement-based quantum computation},
  author={Briegel, Hans J. and Browne, David E. and D{\"u}r, Wolfgang and Raussendorf, Robert and Van den Nest, Maarten},
  journal={Nature Physics},
  volume={5},
  number={1},
  pages={19--26},
  year={2009},
  publisher={Nature Publishing Group UK London}
}

@article{burgholzer2025MQTCore,
	title        = {{{MQT Core}}: {{The}} Backbone of the {{Munich Quantum Toolkit (MQT)}}},
	author       = {Burgholzer, Lukas and Stade, Yannick and Peham, Tom and Wille, Robert},
	year         = 2025,
	journal      = joss,
	publisher    = {The Open Journal},
	volume       = 10,
	number       = 108,
	pages        = 7478,
	doi          = {10.21105/joss.07478},
	url          = {https://doi.org/10.21105/joss.07478}
}

@article{cuccaro2004new,
	title={A new quantum ripple-carry addition circuit},
	author={Cuccaro, Steven A. and Draper, Thomas G. and Kutin, Samuel A. and Moulton, David Petrie},
	journal={arXiv preprint quant-ph/0410184},
	year={2004}
}

@article{deutsch1992rapid,
	title={Rapid solution of problems by quantum computation},
	author={Deutsch, David and Jozsa, Richard},
	journal={Proceedings of the Royal Society of London. Series A: Mathematical and Physical Sciences},
	volume={439},
	number={1907},
	pages={553--558},
	year={1992},
	publisher={The Royal Society London}
}

@article{dobvsivcek2007arbitrary,
	title={Arbitrary accuracy iterative quantum phase estimation algorithm using a single ancillary qubit: {A} two-qubit benchmark},
	author={Dobšíček, Miroslav and Johansson, G{\"o}ran and Shumeiko, Vitaly and Wendin, G{\"o}ran},
	journal=physreviewa,
	volume={76},
	issue={3},
	year={2007},
	month = {9},
	publisher = {American Physical Society},
	doi = {10.1103/PhysRevA.76.030306}
}

@article{draper2000addition,
	title={Addition on a quantum computer},
	author={Draper, Thomas G},
	journal={arXiv preprint quant-ph/0008033},
	year={2000}
}

@article{dur2000three,
	title={Three qubits can be entangled in two inequivalent ways},
	author={D{\"u}r, Wolfgang and Vidal, Guifre and Cirac, J. Ignacio},
	journal=physreviewa,
	volume={62},
	year={2000},
	publisher={APS}
}

@article{farhi2014quantum,
	title={A quantum approximate optimization algorithm},
	author={Farhi, Edward and Goldstone, Jeffrey and Gutmann, Sam},
	journal={arXiv preprint arXiv:1411.4028},
	year={2014}
}

@article{feynman1982simulating,
  title={Simulating physics with computers},
  author={Feynman, Richard P.},
  journal={International Journal of Theoretical Physics},
  volume={21},
  pages={467--488},
  year={1982},
  publisher={Springer},
  doi={https://doi.org/10.1007/BF02650179}
}

@article{fosel2021quantum,
	title={Quantum circuit optimization with deep reinforcement learning},
	author={{F{\"o}sel}, Thomas and {Yuezhen Niu}, Murphy and {Marquardt}, Florian and {Li}, Li},
	journal = {arXiv e-prints},
	year = 2021,
	month = mar,
	eid = {arXiv:2103.07585},
	pages = {arXiv:2103.07585},
	doi = {10.48550/arXiv.2103.07585},
	archivePrefix = {arXiv},
	eprint = {2103.07585},
	primaryClass = {quant-ph}
}

@article{haner2018optimizing,
	title={Optimizing quantum circuits for arithmetic},
	author={H{\"a}ner, Thomas and Roetteler, Martin and Svore, Krysta M},
	journal={arXiv preprint arXiv:1805.12445},
	year={2018}
}

@article{hao2025reducing,
  title={Reducing {T} {G}ates with {U}nitary {S}ynthesis},
  author={Hao, Tianyi and Xu, Amanda and Tannu, Swamit},
  journal={arXiv preprint arXiv:2503.15843},
  year={2026}
}

@article{harrow2009quantum,
	title = {Quantum {A}lgorithm for {L}inear {S}ystems of {E}quations},
	author = {Harrow, Aram W. and Hassidim, Avinatan and Lloyd, Seth},
	journal = {Physical Review Letters},
	volume = {103},
	issue = {15},
	pages = {150502},
	numpages = {4},
	year = {2009},
	publisher = {American Physical Society},
	doi = {10.1103/PhysRevLett.103.150502}
}

@article{karuppasamy2025comprehensive,
	title={A {C}omprehensive {R}eview of {Q}uantum {C}ircuit {O}ptimization: {C}urrent {T}rends and {F}uture {D}irections},
	author={Karuppasamy, Krishnageetha and Puram, Varun and Johnson, Stevens and Thomas, Johnson P.},
	journal={Quantum Reports},
	volume={7},
	number={1},
	pages={2},
	year={2025},
	publisher={MDPI},
	doi={10.3390/quantum7010002}
}

@article{kempe2003quantum,
	title={Quantum random walks: {A}n introductory overview},
	author={Kempe, Julia},
	journal={Contemporary Physics},
	volume={44},
	number={4},
	pages={307--327},
	year={2003},
	publisher={Taylor \& Francis}
}

@article{kitaev1995quantum,
	title={Quantum measurements and the {A}belian stabilizer problem},
	author={Kitaev, A Yu},
	journal={arXiv preprint quant-ph/9511026},
	year={1995}
}

@article{li2024quarl,
	title={Quarl: {A} {L}earning-{B}ased {Q}uantum {C}ircuit {O}ptimizer},
	author={Li, Zikun and Peng, Jinjun and Mei, Yixuan and Lin, Sina and Wu, Yi and Padon, Oded and Jia, Zhihao},
	journal={Proceedings of the ACM on Programming Languages},
	volume={8},
	number={OOPSLA1},
	doi = {10.1145/3649831},
	month = apr,
	articleno = {114},
	numpages = {28},
	year = {2024},
	issue_date = {April 2024},
	publisher = {Association for Computing Machinery}
}

@article{mills2026reinforcement,
  title={Reinforcement {L}earning for {A}daptive {C}omposition of {Q}uantum {C}ircuit {O}ptimisation {P}asses},
  author={Mills, Daniel and Williams, Ifan and Swain, Jacob and Matos, Gabriel and Rinaldi, Enrico and Koziell-Pipe, Alexander},
  journal={arXiv preprint arXiv:2601.21629},
  year={2026}
}

@article{peruzzo2014variational,
	title={A variational eigenvalue solver on a photonic quantum processor},
	author={Peruzzo, Alberto and McClean, Jarrod and Shadbolt, Peter and Yung, Man-Hong and Zhou, Xiao-Qi and Love, Peter J and Aspuru-Guzik, Al{\'a}n and O’brien, Jeremy L},
	journal=natcomms,
	volume={5},
	number={1},
	year={2014},
	publisher={Nature Publishing Group UK London}
}

@article{preskill2013quantum,
  title={Quantum computing and the entanglement frontier},
  author={Preskill, John},
  journal={Bulletin of the American Physical Society},
  volume={58},
  year={2013},
  publisher={APS}
}

@article{quetschlich2023mqtbench,
	title={{{MQT Bench}}: Benchmarking Software and Design Automation Tools for Quantum Computing},
	shorttitle = {{MQT Bench}},
	journal = {{Quantum}},
	author={Quetschlich, Nils and Burgholzer, Lukas and Wille, Robert},
	year={2023},
	note={{{MQT Bench}} is available at \url{https://www.cda.cit.tum.de/mqtbench/}},
}

@article{rattacaso2025quantum,
	title={Quantum circuit compilation with quantum computers},
	author={Rattacaso, Davide and Jaschke, Daniel and Ballarin, Marco and Siloi, Ilaria and Montangero, Simone},
	journal=physreviewresearch,
	volume={7},
	issue={3},
	pages={033268},
	numpages={15},
	year={2025},
	publisher={American Physical Society},
	doi = {10.1103/hcz4-nv2y}
}

@article{rovara2025qubit,
  title={Qubit Reuse Beyond Reorder and Reset: Optimizing Quantum Circuits by Fully Utilizing the Potential of Dynamic Circuits},
  author={Rovara, Damian and Burgholzer, Lukas and Wille, Robert},
  journal={arXiv preprint arXiv:2511.22712},
  year={2025}
}

@article{ruiz2017quantum,
	title={Quantum arithmetic with the quantum {F}ourier transform},
	author={Ruiz-Perez, Lidia and Garcia-Escartin, Juan Carlos},
	journal=qip,
	volume={16},
	number={152},
	year={2017},
	publisher={Springer}
}

@article{vedral1996quantum,
	title={Quantum networks for elementary arithmetic operations},
	author={Vedral, Vlatko and Barenco, Adriano and Ekert, Artur},
	journal=physreviewa,
	volume={54},
	issue={1},
	pages={147--153},
	year={1996},
	publisher = {American Physical Society},
	doi = {10.1103/PhysRevA.54.147}
}

@article{yan2024quantum,
  title={Quantum {C}ircuit {S}ynthesis and {C}ompilation {O}ptimization: {O}verview and {P}rospects},
  author={Yan, Ge and Wu, Wenjie and Chen, Yuheng and Pan, Kaisen and Lu, Xudong and Zhou, Zixiang and Wang, Yuhan and Wang, Ruocheng and Yan, Junchi},
  journal={arXiv preprint arXiv:2407.00736},
  year={2024}
}

@article{ye2026qspe,
  title={{QSPE}: {E}numerating {S}keletal {Q}uantum {P}rograms for {Q}uantum {L}ibrary {T}esting},
  author={Ye, Jiaming and Zhang, Fuyuan and Xia, Shangzhou and Guo, Xiaoyu and Wu, Xiongfei and Zhao, Jianjun and Xue, Yinxing},
  journal={arXiv preprint arXiv:2602.00024},
  year={2026}
}

@article{zen2025reusability,
	title = {Reusability Report: {O}ptimizing {T}-count in General Quantum Circuits with {AlphaTensor-Quantum}},
	author = {Zen, Remmy and N{\"a}gele, Maximilian and Marquardt, Florian},
	journal={Nature Machine Intelligence},
	pages={113--117},
	year={2025},
	publisher={Nature Publishing Group UK London},
	doi={10.1038/s42256-025-01166-9}
}

@book{aho2007compilers,
	title={Compilers: {P}rinciples, {T}echniques and {T}ools},
	author={Aho, Alfred V. and Lam, Monica S. and Sethi, Ravi and Ullman, Jeffrey D.},
	year={2007},
	edition={2},
	isbn={0-321-48681-1},
	publisher={Pearson Education}
}

@book{nielsen2010,
	AUTHOR = {Nielsen, Michael A. AND Chuang, Isaac L.},
	YEAR = {2010},
	TITLE = {Quantum {C}omputation and {Q}uantum {I}nformation - {10th} {A}nniversary {E}dition},
	EDITION = {10},
	ISBN = {978-1-107-00217-3},
	PUBLISHER = {Cambridge University Press}
}

@incollection{greenberger1989going,
	title={Going beyond {B}ell’s theorem},
	author={Greenberger, Daniel M. and Horne, Michael A. and Zeilinger, Anton},
	booktitle={Fundamental Theories of Physics},
	pages={69--72},
	year={1989},
	publisher={Springer Verlag}
}

@inproceedings{arora2025local,
	author={Arora, Jatin and Xu, Mingkuan and Westrick, Sam and Liu, Pengyu and Li, Dantong and Ding, Yongshan and Acar, Umut A.},
	booktitle=qce, 
	title={Local {O}ptimization of {Q}uantum {C}ircuits}, 
	pages={572-583},
	year=2025,
	doi={10.1109/QCE65121.2025.00069}
}

@inproceedings{bernstein1993quantum,
	title={Quantum complexity theory},
	author={Bernstein, Ethan and Vazirani, Umesh},
	booktitle=stoc,
	pages={11--20},
	year={1993}
}

@inproceedings{chen2023quantum,
	title={Quantum constant propagation},
	author={Chen, Yanbin and Stade, Yannick},
	booktitle={International Static Analysis Symposium},
	pages={164--189},
	year={2023},
	organization={Springer}
}

@inproceedings{dang2025qukkos,
	title={Qukkos: {M}easurement-{B}ased {Q}uantum {C}ompilation {S}ystem},
	author={Dang, Son and Son, Youngje and Kim, Brandon},
	booktitle={International Conference on Artificial Intelligence, Computer, Data Sciences and Applications (ACDSA)},
	year={2025},
	pages={1--6},
	organization={IEEE},
	doi={10.1109/ACDSA65407.2025.11165960}
}

@inproceedings{finvzgar2022quark,
	title={QUARK: {A} {F}ramework for {Q}uantum {C}omputing {A}pplication {B}enchmarking},
	author={Fin{\v{z}}gar, Jernej Rudi and Ross, Philipp and H{\"o}lscher, Leonhard and Klepsch, Johannes and Luckow, Andre},
	booktitle=qce,
	pages={226--237},
	year={2022},
	organization={IEEE}
}

@inproceedings{forster2025quantum,
	title={{Q}uantum {C}ircuit {O}ptimization for the {F}ault-{T}olerance {E}ra: {D}o {W}e {H}ave to {S}tart from {S}cratch?},
	author={Forster, Tobias V. and Quetschlich, Nils and Wille, Robert},
	booktitle=qce,
	volume={1},
	pages={584--590},
	year={2025},
	organization={IEEE}
}

@inproceedings{grover1996fast,
	title={A fast quantum mechanical algorithm for database search},
	author={Grover, Lov K.},
	isbn = {0897917855},
	publisher = {{ACM}},
	url = {https://doi.org/10.1145/237814.237866},
	doi = {10.1145/237814.237866},
	booktitle = stoc,
	year={1996},
	pages = {212–219},
	numpages = {8},
}

@inproceedings{hopf2026integrating,
	title={Integrating {Q}uantum {S}oftware {T}ools with (in) {MLIR}},
	author={Hopf, Patrick and Ochoa, Erick and Stade, Yannick and Rovara, Damian and Quetschlich, Nils and Florea, Ioan Albert and Izaac, Josh and Wille, Robert and Burgholzer, Lukas},
	booktitle=sca_hpcasia,
	pages={42--54},
	year={2026}
}

@inproceedings{hopf2026quantum,
	title={{Q}uantum {C}ircuit {C}ompilation for {S}uperconducting {B}us-{R}esonator {A}rchitectures},
	author={Hopf, Patrick and Burgholzer, Lukas and Wille, Robert},
	booktitle=date,
	year={2026}
}

@inproceedings{lattner2021mlir,
	title={MLIR: {S}caling compiler infrastructure for domain specific computation},
	author={Lattner, Chris and Amini, Mehdi and Bondhugula, Uday and Cohen, Albert and Davis, Andy and Pienaar, Jacques and Riddle, River and Shpeisman, Tatiana and Vasilache, Nicolas and Zinenko, Oleksandr},
	booktitle={International Symposium on Code Generation and Optimization},
	year={2021},
	pages={2--14},
	organization={IEEE}
}

@inproceedings{praba2025integration,
	title={Integration of {M}achine {L}earning with {Q}uantum {C}omputing: {C}ompilation of {Q}uantum {C}ircuit for {E}fficacy on {N}oisy {I}ntermediate-{S}cale {Q}uantum {D}evices},
	author={Praba, M. and Samhan, Ahmad Abdelhafiz Ali and Pandi, V. Samuthira and Sowmiya, A. and Ravi, Navyatha and Karthikeyan, G.},
	booktitle={Asian Conference on Innovation in Technology},
	year={2025},
	pages={1--6},
	organization={IEEE},
	doi={10.1109/ASIANCON66527.2025.11281256}
}

@inproceedings{quetschlich2023compiler,
	title={Compiler optimization for quantum computing using reinforcement learning},
	author={Quetschlich, Nils and Burgholzer, Lukas and Wille, Robert},
	booktitle=dac,
	year={2023},
	pages={1--6},
	organization={IEEE},
	doi={10.1109/DAC56929.2023.10248002}
}

@inproceedings{quetschlich2023predicting,
	title={Predicting {G}ood {Q}uantum {C}ircuit {C}ompilation {O}ptions},
	author={Quetschlich, Nils and Burgholzer, Lukas and Wille, Robert},
	booktitle=qsw,
	pages={43--53},
	year={2023},
	organization={IEEE},
	doi={10.1109/QSW59989.2023.00015}
}

@inproceedings{quetschlich2024towards,
	title={Towards {A}pplication-{A}ware {Q}uantum {C}ircuit {C}ompilation},
	author={Quetschlich, Nils and Kiwit, Florian J and Wolf, Maximilian A. and Riofrio, Carlos A. and Burgholzer, Lukas and Luckow, Andre and Wille, Robert},
	booktitle=qsw,
	pages={135--142},
	year={2024},
	organization={IEEE}
}

@inproceedings{remme2025optimization,
	title={Optimization of {H}ybrid {Q}uantum-{C}lassical {A}lgorithms},
	author={Remme, Lian and Weinert, Alexander and Waschk, Andre},
	booktitle=qsw,
    pages={215--226},
    year={2025},
    organization={IEEE},
    doi={10.1109/QSW67625.2025.00033}
}

@inproceedings{shor1994algorithms,
	title={Algorithms for {Q}uantum {C}omputation: {D}iscrete {L}ogarithms and {F}actoring},
	author={Shor, Peter W.},
	booktitle={Annual Symposium on Foundations of Computer Science},
	pages={124--134},
	year={1994},
	organization={{IEEE}}
}

@inproceedings{sunkel2025quantum,
	title={Quantum {C}ircuit {c}onstruction and {O}ptimization through {H}ybrid {E}volutionary {A}lgorithms},
	author={S{\"u}nkel, Leo and Altmann, Philipp and K{\"o}lle, Michael and Stenzel, Gerhard and Gabor, Thomas and Linnhoff-Popien, Claudia},
	booktitle=gecco,
	pages={934--942},
	year={2025},
	doi={10.1145/3712256.3726392}
}

@inproceedings{uotila2025perspectives,
  title={Perspectives on {U}tilization of {M}easurements in {Q}uantum {A}lgorithms},
  author={Uotila, Valter and Salmenper{\"a}, Ilmo and Becker, Leo and Meijer-Van De Griend, Arianne and Shinde, Aakash Ravindra and Nurminen, Jukka K},
  booktitle=qsw,
  year={2025},
  pages={48--59},
  organization={IEEE}
}

@inproceedings{wille2024mqt,
	title        = {The {{MQT}} Handbook: {{A}} Summary of Design Automation Tools and Software for Quantum Computing},
	shorttitle   = {{The MQT Handbook}},
	author       = {Wille, Robert and Berent, Lucas and Forster, Tobias and Kunasaikaran, Jagatheesan and Mato, Kevin and Peham, Tom and Quetschlich, Nils and Rovara, Damian and Sander, Aaron and Schmid, Ludwig and Schoenberger, Daniel and Stade, Yannick and Burgholzer, Lukas},
	year         = 2024,
	booktitle    = qsw,
	doi          = {10.1109/QSW62656.2024.00013},
	eprint       = {2405.17543},
	eprinttype   = {arxiv},
	addendum     = {A live version of this document is available at \url{https://mqt.readthedocs.io}}
}

@online{ibm_aachen_infos,
	author = {IBMQuantum},
	title = {ibm\_aachen},
	year = 2026,
	url = {https://quantum.cloud.ibm.com/computers?system=ibm\_aachen},
	urldate = {2026-02-02}
}

@online{ibm_boston_infos,
	author = {IBMQuantum},
	title = {ibm\_torino},
	year = 2026,
	url = {https://quantum.cloud.ibm.com/computers?system=ibm\_boston},
	urldate = {2026-02-02}
}

@online{ibm_kingston_infos,
	author = {IBMQuantum},
	title = {ibm\_kingston},
	year = 2026,
	url = {https://quantum.cloud.ibm.com/computers?system=ibm\_kingston},
	urldate = {2026-02-02}
}

@online{intel_cpu_infos,
	author = {Intel Corporation},
	title = {{Intel® Core™, Intel® Core Ultra, and Intel Processor Comparison Chart for Laptops}},
	year = 2026,
	url = {https://www.intel.com/content/www/us/en/content-details/843334/intel-core-intel-core-ultra-and-intel-processor-comparison-chart-for-laptops.html},
	date= {2026-01-09},
	urldate = {2025-02-02}
}

@online{ionq_aria,
	author = {IonQ, Inc.},
	title = {{IonQ} {A}ria: {P}ractical {P}erformance},
	year = 2026,
	url = {https://ionq.com/resources/ionq-aria-practical-performance},
	date = {2025-01-08},
	urldate = {2026-02-02}
}

@online{mqt_core_fork,
	title = {{GitHub - LiRem101/mqt-core at mlir/quantum-resource-reduction · GitHub}},
	year = 2026,
	url = {https://github.com/LiRem101/mqt-core/tree/mlir/quantum-resource-reduction},
	urldate = {2026-05-11}
}

@online{qnn_qiskit,
	author = {IBMQuantum},
	title = {EstimatorQNN - {Q}iskit {M}achine {L}earning 0.9.0},
	year = 2026,
	url = {https://qiskit-community.github.io/qiskit-machine-learning/stubs/qiskit\_machine\_learning.neural\_networks.EstimatorQNN.html},
	urldate = {2026-02-27}
}

@online{qiskit_real_amplitudes,
	author = {IBMQuantum},
	title = {{RealAmplitudes} | {IBM} {Q}uantum {D}ocumentation},
	year = 2026,
	url = {https://quantum.cloud.ibm.com/docs/de/api/qiskit/qiskit.circuit.library.RealAmplitudes2},
	urldate = {2026-02-27}
}

@online{qiskit_su_2,
	author = {IBMQuantum},
	title = {{EfficientSU2} | {IBM} {Q}uantum {D}ocumentation},
	year = 2026,
	url = {https://quantum.cloud.ibm.com/docs/de/api/qiskit/qiskit.circuit.library.EfficientSU2},
	urldate = {2026-02-27}
}

@online{qiskit_two_local,
	author = {IBMQuantum},
	title = {{TwoLocal} | {IBM} {Q}uantum {D}ocumentation},
	year = 2026,
	url = {https://quantum.cloud.ibm.com/docs/de/api/qiskit/qiskit.circuit.library.TwoLocal},
	urldate = {2026-02-27}
}

@online{quark_github,
	author = {QUARK-framework},
	title = {GitHub - {QUARK-framework/QUARK}: {F}ramework for {Q}uantum {C}omputing {A}pplication {B}enchmarking (until version 2.2)},
	year = 2025,
	url = {https://github.com/QUARK-framework/QUARK},
	date = {2025-07-30},
	urldate = {2026-02-27}
}
\balance

\end{document}